\documentstyle[prd,aps,epsfig,floats,axodraw]{revtex}
\begin{document}

\draft
\renewcommand{\topfraction}{0.8} 
\newcommand{\beq}{\begin{equation}}
\newcommand{\eeq}{\end{equation}}
\newcommand{\bea}{\begin{eqnarray}}
\newcommand{\eea}{\end{eqnarray}}
\newcommand{\pbar}{\not{\!\partial}}
\newcommand{\dbar}{\not{\!{\!D}}}
\def\lsim{\:\raisebox{-0.75ex}{$\stackrel{\textstyle<}{\sim}$}\:}
\def\gsim{\:\raisebox{-0.75ex}{$\stackrel{\textstyle>}{\sim}$}\:}
\twocolumn[\hsize\textwidth\columnwidth\hsize\csname 
@twocolumnfalse\endcsname

\title{Leptogenesis from a sneutrino condensate revisited}   
\author{Rouzbeh Allahverdi$^{a}$ and Manuel Drees$^{b}$}
\address{$^{a}$Theory Group, TRIUMF, 4004 Wesbrook Mall, Vancouver, B.C., V6T 
2A3, Canada.\\
$^{b}$Physik Department, Technische Universit\"at M\"unchen, James Franck
Strasse, D85748, Garching, Germany.}
\date{\today} 
\maketitle

\begin{abstract}

We re--examine leptogenesis from a right--handed sneutrino condensate,
paying special attention to the $B-$term associated with the see--saw
Majorana mass. This term generates a lepton asymmetry in the
condensate whose time average vanishes. However, a net asymmetry will
result if the sneutrino lifetime is not much longer than the period of
oscillations. Supersymmetry breaking by thermal effects then yields a
lepton asymmetry in the standard model sector after the condensate
decays. We explore different possibilities by taking account of both
the low--energy and Hubble $B-$terms. It will be shown that the
desired baryon asymmetry of the Universe can be obtained for a wide
range of Majorana mass.

\end{abstract}

\vskip2pc]

\section{Introduction}

The baryon asymmetry of the Universe (BAU) parameterized as $\eta_{\rm
B}=(n_{\rm B}-n_{\bar{\rm B}})/s$, with $s$ being the entropy density,
is determined to be $0.9 \times 10^{-10}$, with a precision of $\sim
4\%$, by the recent WMAP data on the temperature anisotropy in cosmic
microwave background (CMB)~\cite{wmap}. This is also in good agreement
with an independent determination from big bang nucleosynthesis
(BBN)~\cite{osw} of the light elements. This asymmetry can be produced
from a baryon symmetric Universe provided three conditions are met:
$B$ and/or $L-$violation, $C-$ and $CP-$violation, and departure from
thermal equilibrium~\cite{sakharov}.  Moreover, $B+L$--violating
sphaleron transitions are active at temperatures $T$ from $10^{12}$
GeV down to $100$ GeV~\cite{krs}. Any mechanism for creating a baryon
asymmetry at $T > 100$ GeV therefore has to create a $B-L$
asymmetry. The final baryon asymmetry is then given by $B=a(B-L)$,
where $a=28/79$ in case of the standard model (SM) and $a=32/92$ for
the minimal supersymmetric standard model (MSSM)~\cite{khlebnikov}.

Leptogenesis is an attractive mechanism for producing a $B-L$
asymmetry~\cite{fy}. This scheme postulates the existence of
right--handed (RH) neutrinos, which are SM singlets, with a lepton
number violating Majorana mass $M_N$. Such a mass is compatible with
all SM symmetries, and hence can be arbitrarily large beyond the
electroweak scale.  This provides an elegant explanation for the small
masses of the light neutrinos via the see--saw
mechanism~\cite{seesaw}. Moreover, a lepton asymmetry can be generated
from the out--of--equilibrium decay of the RH neutrinos to the Higgs
boson and light leptons, provided $CP-$violating phases exist in the
neutrino Yukawa couplings. The lepton asymmetry thus obtained will be
partially converted to a baryon asymmetry via sphaleron effects. This
is the standard lore for leptogenesis~\cite{fy,luty,plumacher}. In
this scenario the on--shell RH neutrinos whose decay is responsible
for the lepton asymmetry can be produced thermally or
non--thermally. In thermal leptogenesis the RH neutrinos are produced
from the primordial thermal bath through their Yukawa
interactions. However, at least one RH neutrino must have small Yukawa
couplings in order to decay sufficiently late, i.e. out of thermal
equilibrium~\cite{plumacher}. The generation of an acceptable lepton
asymmetry then requires the mass $M_1$ of the lightest RH neutrino and
the temperature of the thermal bath to exceed
$10^8$~GeV~\cite{buchmuller,sacha,gnrrs} (unless RH neutrinos are 
degenerate~\cite{pilaftsis}).

However, this is marginally compatible with the upper bound on the
reheat temperature $T_{\rm R}$ in supersymmetric theories, which is
constrained by thermal gravitino production \cite{ellis}. Gravitinos
with mass $m_{3/2}$ of the order of the electroweak scale decay long
after BBN and their decay products can distort the primordial
abundance of the light elements. For $100~{\rm GeV}\lsim m_{3/2} \lsim
1~{\rm TeV}$, a successful nucleosynthesis requires $n_{3/2}/s \leq
(10^{-14}-10^{-12})$, which translates into the bound $T_{\rm R}\leq
(10^{7}-10^{9})~{\rm GeV}$ on the reheat
temperature~\cite{ellis,subir}\footnote{Non--thermal gravitino
production during preheating~\cite{non} does not give rise to any
threat in realistic models of inflation~\cite{abm,nps}. Also, possible
gravitino production from perturbative decays of the
inflaton~\cite{nop}, and/or from heavy long--lived neutral
particles~\cite{aem} will not yield severe bounds.}.

An interesting alternative is non--thermal leptogenesis. In this
scenario RH neutrinos are produced from the decay of the
inflaton~\cite{infl}, and the reheat temperature can be significantly
below $M_N$. This can occur for a perturbative inflaton decay, called
reheating, provided that the RH neutrinos are lighter than the
inflaton~\cite{reheat}. The RH neutrinos can also be produced
non--perturbatively via preheating~\cite{gprt}, or tachyonic
preheating~\cite{tachyon}, even if their mass is larger than the
inflaton mass. Non--thermal leptogenesis can also be achieved without
exciting on--shell RH neutrinos~\cite{bb,am,lazarides}. This allows a
sufficiently low reheat temperature, and can yield the required baryon
asymmetry for a rather wide range of the inflationary scale.

In supersymmetric models one also has the RH sneutrinos which serve as
an additional source for leptogenesis~\cite{cdo}. The sneutrinos are
produced along with neutrinos in a thermal bath or during reheating,
and with much higher abundances in preheating~\cite{bdps}. Moreover,
there are two unique possibilities for leptogenesis from the RH
sneutrinos. First, they can acquire a large vacuum expectation value
(VEV) if their mass during inflation is less than the Hubble expansion
rate at that epoch $H_{I}$. This condensate starts oscillating once $H
\simeq M_N$, thereby automatically satisfying the out--of--equilibrium
condition. The decay of the sneutrino condensate can then yield the
desired lepton asymmetry in the same fashion as neutrino decay
does~\cite{my,hmy}, or~\cite{bmp} via the Affleck--Dine
mechanism~\cite{ad,qball}. The second possibility is to generate the
lepton asymmetry in the RH sneutrino
sector~\cite{adm,nir,agr}\footnote{Note that, due to the Majorana
nature of the RH neutrinos, no lepton asymmetry can be created in that
sector.}. This can be done via inflaton decay to the RH
sneutrinos~\cite{adm}, or from soft supersymmetry breaking
effects~\cite{nir,agr}. This asymmetry will be transferred to the
light (s)leptons upon the decay of the RH sneutrinos, and partially
converted to baryon asymmetry via sphalerons.  However, as emphasized
in~\cite{nir,agr}, the final asymmetry depends on the strength of
supersymmetry breaking effects and will vanish in the supersymmetric
limit.

In this note we re--examine the generation of a lepton asymmetry from a
RH sneutrino condensate in the light of new proposals. We will focus
on the role of the soft supersymmetry breaking $B-$term associated
with the neutrino Majorana mass. This term creates an oscillating
asymmetry in the RH sneutrino condensate whose average, when taken
over many oscillations, vanishes. However, the condensate does carry a
lepton asymmetry at any given time, which can be accessed if the
sneutrino lifetime is not much longer than the oscillation period. We
will consider different cases by taking account of both the low--energy
and Hubble $B-$terms. It will be shown that condensate decay can
result in an acceptable BAU through supersymmetry breaking by
statistics and finite temperature mass corrections. It is important
for the success of this scenario that the sneutrino condensate is not
destroyed before an ${\cal O}(1)$ asymmetry is produced by the
$B-$term(s). We will consider two possible effects in this respect,
namely resonant decay of the condensate and thermal effects. We will
show that thermal effects can prevent resonant decay, while being
sufficiently weak in order not to affect the condensate dynamics
themselves. We will also comment on the possibility that the RH
sneutrino is the inflaton or the curvaton, in which case the sneutrino
condensate dominates the energy density at the time of its decay. Our
main conclusion is that an acceptable baryon asymmetry can be obtained
for wide ranges of the Majorana mass and $B$ term, either
Hubble--induced or from low energy supersymmetry breaking. This
therefore provides a viable alternative for successful leptogenesis
from a sneutrino condensate, which works for a single generation and
does not require any parameter in the Lagrangian to have a nontrivial
phase.

The remainder of this article is organized as follows. In the next
Section we discuss and solve the evolution equation of the sneutrino
condensate in a matter--dominated universe. Sec.~III deals with
perturbative sneutrino decays, with emphasis on supersymmetry breaking
by thermal effects. In Sec.~IV we discuss possible non--perturbative
decays of the $\tilde N$ condensate, and ways to shut them off; we
also derive an upper bound on the reheat temperature from the
requirement that scattering off the thermal bath does not destroy the
coherence of the condensate. Numerical results leading to successful
leptogenesis consistent with all constraints are presented in Sec.~V,
while Sec.~VI contains a discussion of special features of our
mechanism, and briefly sketches the consequences of loosening some of
our assumptions. Finally, Sec.~VII is devoted to a short summary and
some conclusions.

\setcounter{footnote}{0}
\section{Lepton asymmetry in the sneutrino condensate}

We work in the framework of the MSSM augmented with three RH neutrino 
multiplets in order to accommodate neutrino masses via the see--saw 
mechanism~\cite{seesaw}. The relevant part of the superpotential is
\beq \label{superpot}
W \supset {1 \over 2} M_N {\bf N} {\bf N} + h {\bf H}_u {\bf N} {\bf L},
\eeq
where ${\bf N}$, ${\bf H}_u$, and ${\bf L}$ are multiplets containing
the RH neutrinos $N$ and sneutrinos $\tilde N$, the Higgs field giving
mass to the top quark and its superpartner, and the left--handed
(s)lepton doublets, respectively. $h$ are the neutrino Yukawa
couplings and, for simplicity, family indices on $M_N$, $h$, ${\bf
N}$, and ${\bf L}$ are omitted. We work in the basis where the
Majorana mass matrix is diagonal. Note that we are not concerned with
the origin of this mass. It can come from an explicit mass term or
from spontaneous breaking of some symmetry, e.g. in models with a
gauged $U(1)_{B-L}$.

In addition to the supersymmetry conserving part of the scalar potential for 
$\tilde N$, one also has soft terms from 
low--energy supersymmetry breaking~\cite{nilles}
\beq \label{low}
m^2_0 |{\tilde N}|^2 + (B M_N {\tilde N}^2 + {\rm h.c.}),
\eeq
and supersymmetry breaking by the energy density of the Universe
(called Hubble--induced)~\cite{drt1}
\beq \label{hubble}
C_I H^2 |{\tilde N}|^2 + (b H M_N {\tilde N}^2 + {\rm h.c.}).
\eeq
The soft supersymmetry breaking parameters $m_0$ and $B$ typically are
$100~{\rm GeV}-1~{\rm TeV}$ at the weak scale.  In models with
gauge--mediated supersymmetry breaking, $B$ can be very small at the
``messenger'' scale where supersymmetry breaking is transmitted to the
visible sector.  For $|C_I| \sim {\cal O}(1)$ and $|b| \sim {\cal
O}(1)$, Hubble--induced supersymmetry breaking is dominant as long as
$H > m_0$; note that $H \gg m_0$ during inflation. It is known that
all scalar fields with mass less than $H_I$ can acquire a VEV during
inflation, due to the accumulation of quantum
fluctuations~\cite{infl}. Here we consider the case where $M_N <
H_I$\footnote{Note that the RH (s)neutrinos can be massless during
inflation. This happens, e.g., in the model of Ref.~\cite{bmp}, where
the Majorana mass term is generated via the Higgs mechanism after
inflation.}. Therefore the flatness of the $\tilde N$ potential
crucially depends on the sign and size of $C_I$. If $C_I \gsim 1$,
$\tilde N$ will settle at the origin and a condensate will not be
formed. For $0 < C_I \ll 1$, quantum fluctuations will grow along the
$\tilde N$ direction during inflation. These fluctuations can push
$|\langle {\tilde N} \rangle|$ to a maximum value of
$H^{2}_{I}/M_N$~\cite{infl}. Higher VEVs are also possible as initial
condition. On the other hand, if $C_I < 0$, the origin is an unstable
point and higher dimensional terms in the scalar potential set the
minimum of $\tilde N$.

The only renormalizable superpotential term of this type is $\lambda
{\bf N} {\bf N} {\bf N}$. However, this term violates R--parity and, in
consequence, destabilizes the lightest supersymmetric particle
(LSP). The LSP is only a viable dark matter candidate if this term is
much too small to be relevant in our discussion~\cite{aco}. Another
possibility at the renormalizable level will arise if ${\bf N}$ is
charged under some new symmetry at a scale $M > M_N$ (e.g., in
$SO(10)$ GUTs). In this case a $D-$term contribution $\sim g^2
|{\tilde N}|^4$ appears at scales above $M$, with $g$ being a gauge
coupling, leading to $|\langle {\tilde N} \rangle | \lsim M$. It is
also possible to have non--renormalizable superpotential terms. When
${\bf N}$ remains a singlet up to very high scales terms like
$\lambda_n {\bf N}^n/M^{n-3}$, with $n > 3$ and $\lambda_n \sim {\cal
O}(1)$, are allowed where $M$ is $M_{\rm GUT}$ or $M_{\rm
P}$\footnote{$M_{\rm P}= 2.4 \times 10^{18}$ GeV is the reduced Planck
mass.}. However, such terms can be dangerous (if $\lambda_n$ is not
suppressed) as the potential develops other minima at very large
$|\langle {\tilde N}\rangle |$. The sneutrino field may then get
trapped in (one of) these minima if it initially has a larger VEV,
thus violating R--parity and destabilizing the lightest supersymmetric
particle (LSP). If ${\bf N}$ is charged under some gauge group, these
terms can arise only after spontaneous breaking of the new symmetry
typically resulting in $\lambda_n \ll 1$. As an example, consider the
$SO(10)$ GUT. Then $\lambda_n$ has powers of $M_{\rm GUT}/M_{\rm P}$
and/or $M_N/M_{\rm GUT}$. It is therefore expected that in general the
$D-$term contribution limits $|\langle {\tilde N} \rangle |$ from
above, if $\tilde N$ is a non--singlet.

In what follows we consider $\langle {\tilde N} \rangle $ at the end
of inflation to be a free parameter. For $H > M_N$, the absolute value
of the RH sneutrino $|\tilde N|$ rolls down very slowly. It will start
oscillating around the origin, with an initial amplitude $N_0$, when $H
\simeq M_N$.  The low energy and Hubble $B-$terms, from
Eqs.~(\ref{low}) and (\ref{hubble}), generate a potential for the
phase of $\tilde N$\footnote{It is important to notice that for $H
\leq M_N$, non--renormalizable superpotential terms and the
corresponding low energy and Hubble $A-$terms are subdominant to the
terms in Eqs.~(\ref{superpot}), (\ref{low}) and (\ref{hubble}).}. In
general the phase will change if $\tilde N_0$ is displaced from the
minima of this potential; this corresponds to an angular motion in the
complex plane $(\Re {\rm e} \tilde N, \Im {\rm m} \tilde N)$. This is
generally the case since the phase dependence of $V$ is very weak for
$H > M_N$, and hence $\arg \tilde N_0$ will generically be far away
from the minima. The phase motion results in the creation of a lepton
asymmetry in the $\tilde N$ sector, though with a big difference from
the usual Affleck--Dine mechanism~\cite{ad}. There lepton/baryon
number violation is induced by new physics at a scale so high that its
effects will appear only as non--renormalizable interactions once
heavy degrees of freedom are integrated out~\cite{drt2}. On the other
hand, here the RH sneutrinos take part in the dynamics as their mass
is below the inflationary scale. Therefore the lepton number violating
effects, encoded in the $B-$terms, appear as renormalizable
interactions. This has very important consequences for leptogenesis
from an $\tilde N$ condensate.

A non--renormalizable $A-$term quickly becomes irrelevant: it
multiplies a higher power of the relevant scalar field than the mass
term does, and is thus redshifted more rapidly than the mass term by
the expansion of the Universe. Therefore its only role is to trigger
the phase motion of the condensate by providing an initial
``torque''. Once the $A-$term effectively disappears, due to the
redshift, the scalar field VEV will freely ``rotate'' (i.e. its real
and imaginary parts will oscillate with the {\em same} frequency)
resulting in a constant lepton/baryon asymmetry~\cite{ad}. In our
case, on the other hand, the $B-$term is bilinear in $\tilde N$, and
hence is redshifted in exactly the same way as the mass term. This
implies that the $B-$term is always relevant and the motion along the
angular direction will be oscillatory rather than free rotation
(i.e. the real and imaginary parts will oscillate with {\em different}
frequencies). In consequence, the lepton asymmetry created in the
$\tilde N$ condensate is also oscillating coherently and its time
average vanishes. A sizable net asymmetry can be obtained only if the
condensate decay time is comparable to the period of oscillation of
the lepton number carried by $\tilde N$, which is determined by the
size of the $B-$terms.

\setcounter{footnote}{0}

We now turn to a quantitative description of the behavior of $\tilde N
\equiv (\tilde N_R + i \tilde N_I)/\sqrt{2}$, where $\tilde N_{R,I}$
are real scalar fields. As already mentioned, a crucial role is played
by the $B-$term, which we can take to be real (any phase in $B$ can be
absorbed\footnote{A {\em relative} phase between $B$ and $b$ would
modify the evolution of the $\tilde N$ condensate somewhat during the
(rather short) period of time where $|B| \sim |b|H$. However, this
will lead to qualitatively different behavior only in the highly
unlikely case that $\tilde N_0$ is purely real or purely imaginary. We
will therefore not pursue this possibility any further.} by a
re--definition of $\tilde N$). It lifts the mass degeneracy between
${\tilde N}_R$ and ${\tilde N}_I$; allowing both Hubble--induced and
low--energy SUSY breaking, these masses are given by
\bea \label{masses}
m^2_{R,I} &=& M_N^2 + C_I H^2 + m_0^2 \pm 2 M_N (B + b H) \nonumber \\
&\simeq& [M_N \pm (B + b H)]^2,
\eea
where the upper ($+$) sign applies to $m^2_R$, and the second
approximate equality holds for $M_N \gg B + b H, \, \sqrt{|C_I|}
H$. The initial conditions at $H \simeq M_N$ can be chosen as
\beq \label{ini}
{\tilde N}_R = N_0 {\rm cos} \theta_0~;~{\tilde N}_I = N_0 
{\rm sin} \theta_0~;~\frac{d} {dt} (t \tilde N_R) = \frac{d}{dt}
(t\tilde N_I) = 0,  
\eeq
where $\theta_0$ is a phase, which will generically be ${\cal O}(1)$.
The fields then evolve following the standard equations of
motion~\cite{kt}:
\beq \label{eom}
\ddot{\tilde N}_{R,I} + (3 H + \Gamma_N) \dot{\tilde N}_{R.I} +
m^2_{R,I} \tilde N_{R,I} = 0,
\eeq
where $m^2_{R,I}$ are as in Eq.(\ref{masses}), $\Gamma_N$ is the
$\tilde N$ decay width, which we will compute in the next Section, and
the dots denote differentiation with respect to time. Note that
$\tilde N_R$ and $\tilde N_I$ evolve independently of each other.

As mentioned earlier, we assume that the Hubble parameter during
inflation $H_I \gg M_N$. This means that $H \lsim M_N$, the situation
of interest to us, occurs first during the matter--dominated era, when
the inflaton field oscillates coherently. Let us assume for the moment
that $\tilde N$ decays before the inflaton does. Moreover, we assume
that the total energy in the Universe is dominated by the inflaton; in
Sec.~VI F we will comment on scenarios where the sneutrino itself
dominates the energy density. We then only need to solve
Eq.(\ref{eom}) during the matter--dominated era, where we can set $3 H
= 2/t$, $t$ being the time. The late--time behavior of the solution of
this equation can be given analytically:
\bea \label{sol_eom}
\tilde N_{R,I}(t) = \frac{ {\rm e}^{- \Gamma_N (t-t_M)/2 } }
{t} &\big[& A_{R,I} \cos(f_{R,I}(t)) \nonumber \\
&+& B_{R,I} \sin(f_{R,I}(t)) \big]\,.
\eea
The functions $f_{R,I}$ are given by
\beq \label{fri}
f_{R,I}(t) = \omega_{R,I} (t - t_M) \pm \frac {2 b M_N}{3
\omega_{R,I}} \ln \frac {t}{t_M}\,,
\eeq
with
\beq \label{omega}
\omega_{R,I} = \sqrt{ M_N^2 + m_0^2 \pm 2 B M_N - \Gamma_N^2/4} \simeq
M_N \pm B. 
\eeq
Eqs.(\ref{sol_eom})--(\ref{omega}) solve Eq.(\ref{eom}) up to terms of
relative order $C_I/(M_N t)^2, \ b/(M_N t)^2, \ \Gamma_N/(M_N^2
t)$.\footnote{The $1/t$ factor in Eq.(\ref{sol_eom}) explains why we
chose vanishing time derivative for the ``co--moving'' fields $t
\tilde N_{R,I}$ as initial condition (\ref{ini}): only in this case
can $N_0$ be interpreted as initial amplitude of the oscillation of
the $\tilde N$ field. More general boundary conditions give very
similar results so long as the energy stored in the $\tilde N_{R,I}$
fields remains the same.} The coefficients $A_{R,I}$ and $B_{R,I}$ can
be obtained by matching Eq.(\ref{sol_eom}) at $t = t_M$ to a full
(numerical) solution of Eq.(\ref{eom}). If Hubble--induced SUSY
breaking is negligible, i.e. if $|C_I|,\,|b| \ll 1$,
Eq.(\ref{sol_eom}) can be used at all times $t \gsim 1/M_N$; we will
see in Sec.~V that $\Gamma_N < 10^{-6} M_N$ is required, i.e. the solution
(\ref{sol_eom}) of Eq.(\ref{eom}) becomes nearly exact in this case.

The lepton asymmetry in the ${\tilde N}$ sector is generally given by
\bea \label{nasym}
L_{\tilde N} &\equiv& n_{\tilde N} - n_{{\tilde N}^*} =  i({\dot
{{\tilde N}^*}} {\tilde N} - {\dot {\tilde N}} {\tilde N}^*) \nonumber\\ 
&=&  \dot{\tilde{N}}_I \tilde N_R - \dot{\tilde{N}}_R
\tilde N_I \, .
\eea
Whenever $\tilde N$ is described by Eq.(\ref{sol_eom}), the asymmetry
in the $\tilde N$ condensate amounts to
\bea \label{ln1}
L_{\tilde N} \simeq \frac { {\rm e}^{-\Gamma_N(t-t_M)} } {t^2} &\Big\{
&\omega \big[ (A_R B_I - B_R A_I) \cos(f_R-f_I) \nonumber \\
&& \ \ + (A_R A_I + B_R B_I) \sin(f_R - f_I) \big] \nonumber \\
&-& \Delta \big[ (A_R B_I + A_I B_R) \cos(f_R+f_I)  \\
&& \ \ + (B_R B_I - A_R A_I) \sin(f_R+f_I) \big] \Big\}\,, \nonumber
\eea
with 
\beq \label{delta}
\Delta(t) = \delta + \frac { 2 b M_N} {3 \omega t} \, .
\eeq
Here we have used the notation $\omega_{R,I} = \omega \pm \delta$.  As
mentioned earlier, if Hubble--induced SUSY breaking is negligible, we
can match the solution (\ref{sol_eom}) directly to the initial
condition (\ref{ini}). Ignoring the small, ${\cal O}(B)$ terms in the
last two lines of Eq.(\ref{ln1}), the lepton asymmetry of the
condensate then becomes
\beq \label{ln2}
L_{\tilde N} \simeq \frac{ {\rm e}^{-\Gamma_N t}} {t^2}
M_N N_0^2 ~ \sin2 \theta_0 ~ {\rm sin}(2 B t).
\eeq

The factor $1/t^2$ in Eqs.(\ref{ln1}), (\ref{ln2}) just describes the
redshifting of the ${\tilde N}$ number density; it will drop out when
we normalize the asymmetry to the inflaton or entropy density. Apart
from this factor, the asymmetry in Eq.(\ref{ln2}) reaches its maximum
at $t \simeq (|B|)^{-1}$ but, due to its oscillation with a frequency
$2|B|$, its average vanishes over longer times. The RH sneutrino must
therefore have a decay rate $\Gamma_N \simeq |B|$ in order to transfer
the maximum lepton asymmetry to left--handed (s)leptons. This is a
crucial point to which we will come back in the next section. The
difference from the usual lepton/baryon asymmetry generation from a
condensate is evident. A dimension three or higher $A-$term generates
an asymmetry and becomes irrelevant soon thereafter, thus the
(co--moving) asymmetry approaches a constant. In our case the
asymmetry is generated by a dimension two $B-$term and has an
oscillatory nature. The maximum asymmetry is obtained after a quarter
of oscillation and can be transferred efficiently to the light
(s)leptons only if ${\tilde N}$ condensate decays around the same
time.
                                    
Eq.(\ref{ln1}) shows that the behavior of the lepton asymmetry stored
in the $\tilde N$ condensate becomes more complicated in the presence
of a Hubble--induced $B-$term, since then the difference $f_R - f_L$
contains a term $\propto b \ln \frac {t} {t_M}$. If this term
dominates, it leads to a period of oscillation that grows
exponentially with time. If $b$ is ${\cal O}(1)$, this effectively
sweeps a wide range of oscillation frequencies, and therefore allows
efficient transfer of the asymmetry to light (s)lepton fields for any
$\Gamma_N \in [|B|, M_N {\em e}^{-1/|b|}]$. The size of $|b|$ depends
on the details of the inflationary model and the inflaton coupling to
the RH (s)neutrinos. Since $H \sim m_\phi {\hat \phi}/ M_{\rm P}$,
with $m_\phi$ being the inflaton mass and $\hat \phi$ the amplitude of
inflaton oscillations, $|b|$ is linear in the inflaton field during
the the matter--dominated era of inflaton oscillations. It should
therefore vanish if forbidden by some symmetry (e.g., an R--symmetry),
so long as that symmetry is unbroken. If the inflaton couples to the
RH (s)neutrinos only gravitationally, one generally has $|b| \propto
\phi_0/M_{\rm P}$, where $\phi_0$ is the inflaton VEV at the minimum
of its potential~\cite{aem}. Thus it is expected that $|b| \ll 1$ in
models of chaotic inflation where $\phi_0 = 0$, and new inflation
where $\phi_0 \ll M_{\rm P}$ is possible. In such cases the Hubble
$B-$term is probably negligible, otherwise it should be taken into
account.

One comment is in order before moving on to the next section. Here we
have assumed that $H_u$ and $\tilde L$ have a VEV $\ll N_0$ at the end
of inflation. Otherwise the situation will be considerably more
complicated due to the coupling of these fields to the sneutrino,
leading to the scenario studied in~\cite{sy}. A natural way to achieve
this is to make $H_u$ and $\tilde L$ sufficiently heavy by requiring $h N_0
> H_I$. For an acceptable choice of $h$ and $N_0$ allowing successful
baryogenesis (see next sections), a constraint on the scale of
inflation will thus be obtained. It can be evaded if, for example,
$H_u$ and $\tilde L$ have a Hubble-induced ${\rm mass}^2$ $> H^2$. This
requires the soft mass of $\tilde N$ to be considerably different from
that of $H_u$ and $\tilde L$ at scales of order $H_I$.\footnote{Even
if these masses are comparable at a renormalization scale near $M_{\rm
P}$, they could differ significantly at scale $H_I$ \cite{ad_rad}. A
sufficiently large mass for the $H_u L$ flat direction could for
example arise in the presence of sizable Hubble--induced gaugino
masses and/or Hubble--induced $\mu$ term.}


\section{Sneutrino decay and supersymmetry breaking}
\setcounter{footnote}{0}

It is crucial to transfer the asymmetry generated in the $\tilde N$
condensate to left--handed (LH) (s)leptons, so that it will be
partially converted into a baryon asymmetry by sphalerons. In general
one can imagine two possibilities for creating a lepton/baryon
asymmetry from a condensate. First, the asymmetry is produced in a
condensate carrying lepton/baryon number and then transferred to
fermions via lepton/baryon number conserving interactions. This is
what happens in the Affleck--Dine mechanism~\cite{ad}. The second
possibility is that the condensate does not carry any lepton/baryon
number but some other $U(1)$ charge. In this case the free rotation of
the condensate amounts to an asymmetry in the $U(1)$ charge.  This can
be converted into lepton/baryon asymmetry provided the condensate
decay to fermions violates lepton/baryon number. This possibility has
been considered for baryogenesis from the decay of a complex inflaton
field~\cite{nr}, and baryogenesis in large extra dimension
models~\cite{aemp}. On the other hand, the RH sneutrino carries lepton
number and its decay to light (s)leptons violates lepton number
also. One should therefore examine whether the generated asymmetry in
the condensate survives after the $\tilde N$ decay.

The leading decay channels for $\tilde N$, read from
Eq.(\ref{superpot}), are $\tilde N \rightarrow H_u {\tilde L}$ and
$\tilde N \rightarrow {\bar {\tilde H}_u} {\bar L}$, creating lepton
numbers $+1$ and $-1$ respectively. In the limit of unbroken
supersymmetry both channels have the same decay width $(h^2/8 \pi)
M_N$ (ignoring final state masses, which should be $\ll M_N$ in the
supersymmetric limit). Thus no net asymmetry will be generated among
the left--handed (s)leptons, even for maximal asymmetry in the $\tilde
N$ sector.\footnote{One might wonder that an asymmetry is produced in
the three--body decay channel derived from the four--point vertex
${\tilde N} {\tilde L} {\bar {\tilde {Q}_3}} {{\tilde t}}$, where
$\tilde {Q}_{3}$ and $\tilde t$ are the top squark doublet and singlet
respectively. However, there are three other three--body channels
mediated by the Higgs boson and Higgsino. Once again, in the
supersymmetric limit, the sum of all three--body channels yields no
net asymmetry~\cite{nir}.} This is intuitively understandable as no
lepton asymmetry can be produced from the the decay of RH neutrinos
(in the absence of CP--violating phases). In the supersymmetric limit
one expects that the same also holds for RH sneutrinos.

Supersymmetry breaking is therefore needed for obtaining an asymmetry
in $\tilde N$ decay. The low--energy soft terms result in different
partial widths for $\tilde N$ decays into channels with bosonic and
fermionic final states. The presence of $A-$terms allows the
conjugate two--body decay mode $\tilde N \rightarrow {\bar H}_u {\bar
{\tilde L}}$. Also, due to soft masses, the rates for $\tilde N
\rightarrow {\bar {\tilde H}_u} {\bar L}$ and $\tilde N \rightarrow
H_u {\tilde L}$ channels will be different by an amount of $\sim 2
(m_0/M_N)^2$. The low--energy soft terms therefore yield a lepton
asymmetry suppressed by a factor $(m_0/M_N)^2$ with respect to the
initial $\tilde N$ asymmetry. This is usually too small since $m_0 \ll
M_N$, and hence low--energy supersymmetry breaking alone is not
sufficient for successful leptogenesis.

However, supersymmetry is strongly broken in the early Universe. The
inflaton energy density, as mentioned earlier, induces a soft ${\rm
mass}^2$ of order $H^2$ for the scalars. Also, finite temperature
effects in the primordial thermal bath give rise to a split in the
${\rm mass}^2$ of scalars and fermions $\propto T^2$. We are
interested in supersymmetry breaking at the time of ${\tilde N}$
decay, i.e. when $H \simeq \Gamma_N \ll M_N$.  At this time
Hubble--induced supersymmetry breaking is already quite small. On the
other hand, thermal (supersymmetry breaking) mass corrections can be
substantial since $T \simeq M_N$ is possible. As mentioned earlier, we
assume that the Universe is still matter--dominated at the time of
$\tilde N$ decay. However, for $H > \Gamma_\phi$ the exponential decay
of the inflaton results in a thermal bath with temperature $T = 1.4 (H
T^2_{\rm R} M_{\rm P} g_*^{-1/2})^{1/4}$~\cite{kt}; here $\Gamma_\phi$
is the inflaton decay width, $T_{\rm R}$ is the reheat temperature
(i.e. the temperature of the Universe at the transition from the
matter--dominated to the radiation--dominated epoch, after most
inflatons have decayed), and $g_* \simeq 225$ is the number of light
degrees of freedom in the MSSM. The maximum temperature of this
instantaneous thermal bath $T_{max}$ depends on the thermalization
rate of inflaton decay products (for details on thermalization,
see~\cite{thermalization}). However, we generally expect that $T \gsim
M_N$ is possible if $T_{\rm R} \gsim M_N/50$.

The thermal masses for the slepton $\tilde L$, Higgs $H_u$, lepton
$L$ and Higgsino ${\tilde H}_u$ are~\cite{agr}:
\bea \label{thermal1}
m^2_{H_u} = 2 m^2_{{\tilde H}_u} = \left({3 \over 8} g^2_2 + {1 \over 8} 
g^2_1 + 
{3 \over 4} h^2_t \right) T^2 \nonumber\\ 
m^2_{\tilde L} = 2 m^2_L = \left({3 \over 8} g^2_2 + {1 \over 8} g^2_1 
\right) T^2,
\eea
where $g_1$ and $g_2$ are the $U(1)_Y$ and $SU(2)$ gauge couplings,
respectively, and $h_t$ is the top Yukawa coupling. Thermal
corrections to the sneutrino mass are $\propto h$, thus
negligible. For a detailed study of finite temperature effects on
leptogenesis see~\cite{gnrrs}. At energies of order $M_N$ we take $(3
g_2^2 + g_1^2)/8 = 1/6, \ h_t^2 = 1/2$. We therefore have
\beq \label{thermal2}
m^2_B \equiv m^2_{H_u} + m^2_{\tilde L} \simeq 0.71 T^2~;
~m^2_F \equiv m^2_{{\tilde H}_u} + m^2_L \simeq 0.35 T^2\, .
\eeq
The precise values of $m_{B,F}$ are not very important; however, the
relation $m_B^2 = 2 m_F^2$, which holds independently of the numerical
values of the coupling constants, means that fermionic $\tilde N$
decays open up first.

A second effect of the thermal bath is due to different statistics of
bosons and fermions. Decays into fermionic [bosonic] final states are
suppressed [enhanced] by a factor $(1 - f_F)^2$ [$(1 + f_B)^2$]. Here,
$f_F$ and $f_B$ are the fermion and boson occupation numbers, given by
\beq \label{fermbos}
f_F(E) = \frac {1} { {\rm e}^{E/T} + 1}\, ; \ 
f_B(E) = \frac {1} { {\rm e}^{E/T} - 1}\, .
\eeq

Altogether the total $\tilde N$ decay width in the presence of a
thermal bath can thus be written as
\beq \label{totalrate}
\Gamma_N = \Gamma^B_{N} + \Gamma^F_{N} \, .
\eeq
The partial width into the bosonic $H_u \tilde L$ final state is given
by
\beq \label{gammab}
\Gamma_N^B = \frac {h^2 M_N} {8 \pi} \lambda^{1/2} \left(1, \frac
{m^2_{H_u}} {M_N^2}, \frac {m^2_{\tilde L}} {M_N^2} \right) ( 1 +
f_B)^2 \, ,
\eeq
while the partial width into the fermionic $\bar{\tilde H_u} \bar{L}$
final state is
\beq \label{gammaf}
\Gamma_N^F = \frac {h^2 M_N} {8 \pi} \lambda^{1/2} \left(1, \frac
{m^2 _{\tilde H_u}} {M_N^2}, \frac {m^2_L} {M_N^2} \right) 
( 1 - f_F)^2 \, .
\eeq
Here, the kinematical factor is given by
\beq \label{lambda}
\lambda^{1/2}(1,a,b) = \sqrt{ (1-a-b)^2 - 4ab }\, ,
\eeq
and the functions $f_B, \ f_F$ in Eqs.(\ref{gammab}) and
(\ref{gammaf}) should be taken at $E = M_N/2$. Recall that the
bosonic (fermionic) final state carries lepton number +1 ($-1$). The
efficiency with which a lepton asymmetry can be transferred from the
sneutrino condensate to light (s)leptons is thus described by the
quantity
\beq \label{deltabf}
\Delta_{BF} \equiv \frac {\Gamma_N^B - \Gamma_N^F} {\Gamma_N^B +
\Gamma_N^F}\, .
\eeq
The evolution of the lepton number density carried by the light
$\tilde N$ decay products, $L_D$, is therefore described by the
first--order differential equation
\beq \label{eom_l}
\dot{L}_D + 3 H L_D = L_{\tilde N} \Gamma_N \Delta_{BF}\,,
\eeq
where $L_{\tilde N}$ has been defined in Eq.(\ref{nasym}).

Let us discuss the behavior of $\Delta_{BF}$. At $T > 1.2 M_N$,
thermal effects block all $\tilde N$ decays, i.e. $\Gamma_N =
0$. Eqs.(\ref{thermal2}) show that the leptonic mode opens up first,
i.e. for $1.2 M_N > T > 0.9 M_N$, we have $\Gamma_N = \Gamma_N^F$,
which implies $\Delta_{BF} = -1$. For $T < 0.9 M_N$, the bosonic mode
also becomes allowed, and quickly takes over owing to the rather
strong statistical effects at $T \sim M_N$, see
Eqs.(\ref{fermbos}). At somewhat lower temperatures an expansion in
$T/M_N$ becomes possible; one finds
\beq \label{deltabfapp}
\Delta_{BF} \simeq -{0.18 T^2 \over M^2_N} + 2~\exp\left(-{M_N \over 2
T}\right). 
\eeq
The second term is dominant when $T > M_N/15$ but drops very quickly
for smaller $T$, where the first term takes over, i.e. $\Delta_{BF}$
changes sign for a second time; however, at this point $|\Delta_{BF}|
\lsim 10^{-3}$ is already too small for an efficient transfer of the
lepton asymmetry from the $\tilde N$ condensate to the $\tilde N$
decay products.

In order to find the final baryon asymmetry we again focus on the case
$\Gamma_N > \Gamma_\phi$. In fact, this will be necessary for
successful leptogenesis when $|B| \simeq m_0$. The reason is that
$\Gamma_\phi > \Gamma_N \simeq m_0$ will result in $T_{\rm R} \sim
\sqrt{\Gamma_\phi M_{\rm P}} > 10^{10}$ GeV, thus leading to gravitino
overproduction\footnote{If $\tilde N$ dominates the energy density of
the Universe, its decay can dilute the excess of gravitinos. In that
case $\tilde N$ can act as the curvaton and be responsible for
cosmological density perturbations~\cite{hmy,mm,m}. We will come back to
his possibility in Sec.~VI F.}. At $H = M_N$, we have $\rho_\phi = 3
M^2_N M^2_{\rm P}$. For $\Gamma_\phi < H < m_\phi$ the Universe is
matter--dominated and the inflaton number density is redshifted
$\propto R^{-3}$, where $R \propto t^{2/3}$ is the scale factor of the
Universe; this is nothing but the $1/t^2$ factor of Eqs.(\ref{ln1})
and (\ref{ln2}):
\beq \label{nphi}
n_\phi(t) = \frac {\rho_\phi} {m_\phi} = \frac {3 H^2 M_{\rm P}^2}
{m_\phi} = \frac {4 M_{\rm P}^2} {3 t^2 m_\phi}\,.
\eeq
Eventually the inflatons will decay, thereby reheating the universe,
i.e. releasing a large amount of entropy:
\beq \label{iratio}
{n_\phi(t = \Gamma_\phi^{-1}) \over s} \simeq {3 T_{\rm R} \over 4 m_\phi},
\eeq
where $s$ is the entropy density. Note that for $\Gamma_\phi < H <
\Gamma_N$ the ratio $L_D/n_\phi$ will remain constant. We saw at the
beginning of Sec.~I that the lepton to baryon conversion efficiency
through sphaleron effects is about 1/3. Altogether we thus have as
final asymmetry
\beq \label{final}
\frac {n_B}{s} \simeq \frac{1}{3} \cdot \frac {L_D(t)} {n_\phi(t)} \cdot
\frac{n_\phi(t=\Gamma_\phi^{-1})} {s}\,,
\eeq
where the quantities in the second factor can be taken at any time $t
\gg \Gamma_N^{-1}$. Note that the inflaton mass $m_\phi$ cancels
between the second and the third factor in Eq.(\ref{final}).

In general $L_D$ in Eq.(\ref{final}) has to be computed by numerically
solving Eqs.(\ref{eom}) and (\ref{eom_l}). However, in order to get a
rough estimate of the final asymmetry we can use the approximate
result (\ref{ln2}) for $L_{\tilde N}$, and assume that it is
transferred to $L_D$ with efficiency $\Delta_{BF}$. Moreover, we can
estimate the initial energy density in the sneutrino condensate as
$\rho_{\tilde N} \simeq M_N^2 N_0^2$. For $m_\phi > H > \Gamma_N$ we
then have
\beq \label{nratio}
\frac {n_{\tilde N}} {n_\phi} \simeq \frac {m_\phi N_0^2} {3 M_N
M^2_{\rm P}} \, .
\eeq
Together with Eq.(\ref{iratio}) and again including the 1/3 efficiency
for lepton to baryon conversion, this leads to the estimate
\beq \label{final_app}
\frac{|n_B|} {s} \lsim \frac {T_{\rm R} N_0^2} {12 M_N M^2_{\rm P}}
|\Delta_{BF}| |\sin 2 \theta_0| \,.
\eeq
Here $\Delta_{BF}$ should be computed at temperature $T \simeq
(g_*^{-1/2} \Gamma_N T^2_{\rm R} M_{\rm P})^{1/4}$, which is the
temperature of the thermal plasma at $H = \Gamma_N$. The (approximate)
equality holds if (most) ${\tilde N}$ decays occur while $|t^2
L_{\tilde N}|$ is near its maximum; we saw earlier that this requires
$\Gamma_N \simeq |B|$ if the Hubble--induced contribution to the
$B-$parameter is negligible.

\section{Constraints}
\setcounter{footnote}{0}

So far we have assumed that the sneutrino condensate starts
oscillating when $H \simeq M_N$, and survives until $H \simeq
\Gamma_N$, at which point it decays via perturbative one--particle
decays. However, neither the survival of the condensate as a coherent
state\footnote{We do not expect the condensate to fragment into
$Q-$balls \cite{qball}. This could happen if the potential increased
slower than quadratically at large field values, e.g. if RG evolution
decreased the relevant mass. However, in our case the RG evolution of
$M_N$ is negligible, since $N$ is a gauge singlet with very small
Yukawa couplings. Nevertheless, the condensate could be destroyed by
thermal effects.}, nor its perturbative decay is in general
guaranteed. Here we show that these requirements can be translated
into an upper and lower bound on the reheat temperature $T_{\rm R}$,
respectively. These constraints apply equally to the scenario of
refs.\cite{my,hmy}, which also rely on the perturbative decay of an
$\tilde N$ condensate.

\subsection{Resonant decay of the condensate}

The decay of coherent oscillations of a massive scalar field in cosmology was
initially considered in the context of inflation~\cite{reheating}. The 
energy--momentum tensor of such an
oscillating field, when averaged over oscillations, resembles that of
non--relativistic matter with the same mass.  If the decay occurs over
many oscillations, which is presumably the case for sufficiently weak
couplings, its rate should be the same as the one--particle decay rate
of non--relativistic matter. However, the situation can be more
complicated for larger couplings. Due to the coherent nature of
oscillations, the occupation number of decay products can become
large in which case the one--particle decay approximation
will not be adequate. This was noticed rather recently in the
case of inflaton decay~\cite{preheat1,preheat2}. In fact, depending on the
amplitude of oscillations and the size of its coupling, the condensate
decay can occur in different regimes (typically non--perturbatively)
with different outcomes. Recently, the possibility of non--perturbative
decay of an oscillating $\tilde N$ condensate has also been
discussed~\cite{pm}. Here we briefly go through different regimes for
$\tilde N$ decay, point out implications of a non--perturbative
decay for the lepton asymmetry, and discuss effects which can shut
it off.

We shall focus on the decay of the $\tilde N$ condensate to scalars as it
provides the most efficient channel for energy transfer. The dominant
terms responsible for such decays are derived from the superpotential
in (\ref{superpot})
\beq \label{scalpot}
h^2 |\tilde N|^2 \left[ |H_u|^2 + |\tilde L|^2 \right] + (h M_N 
{\tilde N}^{*} H_u {\tilde L} + {\rm h.c.}).
\eeq
First consider the case when $\hat N \gg M_N/h$, $\hat N$ being
the amplitude of $\tilde N$ oscillations; initially we have $\hat N(H
= M_N) = N_0$, see Eq.(\ref{ini}). In this regime the first two terms
in Eq.~(\ref{scalpot}) are dominant leading to explosive production of $H_u$ 
and $\tilde L$ quanta with (physical) energies $\lsim (h \hat N M_N)^{1/2} 
\gg M_N$. This process, called preheating, typically completes in a time 
scale of several tens of $M^{-1}_N$~\cite{preheat1}. Eventually, due to 
re-scatterings, one has a 
plasma of $\tilde N$, $H_u$ and $\tilde L$ quanta with large occupation 
numbers and typical momenta $\lsim (h {\hat N} M_N)^{1/2}$. The preheat 
plasma (and every other field which is coupled to it) is in kinetic 
equilibrium, but full thermal equilibrium takes much longer to 
establish~\cite{fk}. It is noticeable that particles with a mass up to $(h 
{\hat N} M_N)^{1/2}$ can be produced during preheating.

The amplitude $\hat N$ is redshifted $\propto 1/t$ if preheating is
blocked [see Eq.(\ref{nopreheat}) below]. Eventually we have $\hat N
\ll M_N/h$, where the cubic terms in Eq.~(\ref{scalpot}) will be
dominant. If $M^2_N/(h^2 N_0) \ll \hat N \ll M_N /h$, the condensate
decay occurs in the narrow--band resonance
regime~\cite{preheat1,preheat2}; note that the first of these strong
inequalities requires $N_0 \gg M_N/h$, since $\hat N \leq N_0$. In
this regime $H_u$ and $\tilde L$ quanta with occupation numbers larger
than one and energy narrowly peaked around $M_N/2$ are produced, at a
rate given in~\cite{preheat2}. If $\hat N \ll M^2_N/(h^2 N_0)$, the
decay is perturbative and its rate will be given by the familiar
one--particle decay rate. Finally, if $h \hat N \simeq M_N$, all terms
in (\ref{scalpot}) are comparable. In this case the cubic terms result
in a tachyonic instability during part of oscillations which can again
lead to a rapid decay of $\tilde N$~\cite{pm}.

The decay of $\tilde N$ oscillations via preheating has undesirable
consequences for successful leptogenesis. First, $\Gamma_{\rm preheat}
\gg |B|$, and hence the asymmetry in the condensate, given in
Eq.~(\ref{nasym}), is suppressed. Moreover, since particles in the
preheat plasma have energies $\gg M_N$, lepton number violating
processes mediated by $\tilde N$ and $N$ will be efficient. Therefore
any existing lepton asymmetry will quickly be washed
away\footnote{Since preheating destroys the coherence of $\tilde N$
oscillations, the $B-$term will not generate a (large) net asymmetry from the
preheat plasma.}. The situation is somewhat better for narrow--band
resonance decay. There are two effects competing with each other in
this case. On the one hand, $\Gamma_{\rm narrow} > |B|$ which
suppresses the asymmetry in the condensate unless the Hubble $B$-term is 
sizeable. On the other hand,
$\Delta_{BF} \simeq 1$ as the condensate essentially decays to
scalars. Therefore the yielded lepton asymmetry can actually be larger
than that in a perturbative decay, unless $\Gamma_{\rm narrow} \gg
|B|$.

The picture given above is for a simple toy model. In realistic models
resonant decay can be altered, suppressed or postponed due to various
effects. For example, final state self--couplings of moderate strength
can qualitatively alter resonant decay~\cite{ac,pr}. Also, for a
complex oscillating field, preheating can be postponed due to
out--of--phase oscillations of the real and imaginary
components~\cite{acs}. In the case of sneutrino both of these effects
are present. The decay products $H_u$ and $\tilde L$ have $D-$term
couplings of gauge strength to each other, and the $B-$term causes
out--of--phase oscillations of ${\tilde N}_R$ and ${\tilde N}_I$. These,
as pointed out in~\cite{pm}, can regulate the resonant decay of
$\tilde N$ oscillations.  

In the absence of a complete understanding of possible
non--perturbative decays of the $\tilde N$ condensate, it is safer to
require that these non--perturbative effects are suppressed.  One
possibility is to require that $h N_0 < M_N$. However, we will see in
Sec.~V that this results in too small a baryon asymmetry. The second
possibility is to prevent resonant decays until $\hat N$ is redshifted
to a sufficiently small value. This will in fact happen if the medium
is sufficiently hot.  We saw that preheating can produce particles
with mass up to $(h N_0 M_N)^{1/2}$. Together with
Eqs.(\ref{thermal2}) this implies that thermal effects can
kinematically shut off preheating if $T_0^2 > h N_0 M_N$, which
requires
\beq \label{nopreheat}
T^2_{\rm R} M_{\rm P} > \frac {\sqrt{g_*}} {4} h^2 N^2_0 M_N\,.
\eeq
Since (in the absence of preheating) the oscillation amplitude scales
like $1/t$, while $T^2 \propto 1/\sqrt{t}$, the constraint
(\ref{nopreheat}) also forbids non--perturbative $\tilde N$ decay,
including decays in the narrow--band resonance regime, at all later
times. It is therefore a sufficient condition for ensuring that the
sneutrino condensate decays perturbatively.

We want these perturbative decays to occur when $H \simeq h^2 M_N /
(4\pi)$. We saw in the discussion of Eq.(\ref{eom_l}) that this
requires $T < 1.2 M_N$ at that time, which implies
\beq \label{pert}
\frac{h^2}{4\pi} T^2_{\rm R} M_{\rm P} < M^3_N \sqrt{g_*}.
\eeq
Otherwise thermal corrections postpone the perturbative
decay~\cite{ac,knr}. This may actually be desired, if $|B| < h^2 M_N /
(4\pi)$. However, we will see below that the constraint (\ref{pert})
has to be satisfied if {\em coherent} oscillations of the condensate
are to continue until $H \simeq \Gamma_N$.

\subsection{Thermal destruction of the condensate}

One has to also ensure that the coherence of oscillations is not lost
by scatterings off particles in the thermal bath~\cite{thermal}. The
most efficient reactions are $\tilde N$ annihilation with $H_u$ and
${\tilde H}_u$, $L$ and $\tilde L$, top (s)quarks and right-handed
bottom (s)quarks; the corresponding matrix elements are $\propto h
h_t$\footnote{There are also annihilation diagrams including
electroweak gauge and gaugino fields. These have a smaller
multiplicity factor, and are hence subdominant. In addition, $2
\rightarrow 1$ annihilations with (s)leptons into an on--shell $H_u$
or $\tilde H_u$, which have a larger phase space factor, can also be
active. These, however, are kinematically allowed when $T$ is
sufficiently higher than $M_N$. Therefore, since $H \propto T^4$, they
can be safely neglected.}. Coherence will be lost if the rate for such
reactions is faster than the expansion rate $H$. We therefore have to
require
\beq \label{condsurv}
\frac{63 \xi(3) h^2 h_t^2} {16 \pi^3} \cdot \frac{T^3} {M_N^2 + 6 T
M_N} < H\,.
\eeq
The factor 63 counts the final state multiplicities and number of
initial state degrees of freedom (which enter the target density); the
product of these two coefficients is in fact identical for all bosonic
reactions. We have included a factor 1.75 to take annihilation on a
fermion into account. The factor $T^3$ also comes from the density of
particles in the plasma, and $M_N^2 + 6 T M_N$ is a typical squared
center--of--mass energy (at $T < M_N$). The condition (\ref{condsurv})
has to be satisfied at all times between the onset of oscillations ($H
\simeq M_N$) until their perturbative decay [$H \simeq h^2 M_N / (4
\pi)$]. In the matter--dominated era, $H \propto T^4$. Condition
(\ref{condsurv}) is therefore most difficult to satisfy at the {\em
latest} relevant time. Conservatively ignoring the $6 T M_N$ term in
the denominator, and setting $h_t^2 = 0.5$ and $T = 1.4 \left( T_R^2 H
M_P g_*^{-1/2} \right)^{1/4}$ as in Sec.~II, we find as relevant
constraint at $H = h^2 M_N/(4\pi)$:
\beq \label{trmax}
\frac{h^2} {4\pi} T_{\rm R}^2 M_P < \frac{\sqrt{g_*}}{4} M_N^3\,.
\eeq
This bound is very similar to, although slightly stronger than, the
condition (\ref{pert}) above, i.e. it implies that thermal effects do
not delay the perturbative decay of the $\tilde N$ condensate
significantly.

Moreover, it is also possible that thermal effects trigger early
oscillations of $\tilde N$~\cite{thermal}. This will happen, provided
that $h N_0 < T$ (so that Higgs/Higgsino and (s)leptons will be in
thermal equilibrium) and $hT > H$ (so that they induced a thermal mass
for $\tilde N$ exceeding the Hubble rate) when $H > M_N$. Early
oscillations dampen $\tilde N$ more than the case with $H_{osc} \simeq
M_N$, thus leading to a smaller lepton asymmetry. Also, so long as $hT
> M_N$, the thermal mass of the sneutrino $hT$ replaces $M_N$ in all
calculations. However, $hT < M_N$ at the time of sneutrino decay since
otherwise the large thermal mass of final states kinematically blocks
the decay. Therefore the condition for preventing evaporation of the
condensate is still given by~(\ref{trmax}). This condition implies
that $hT > M_N$ is only possible at $H > M_N / (4\pi h^2)$. If the
constraint in~(\ref{trmax}) is saturated , $hT = M_N$ implies $H=
M_N/(4 \pi h^2)$; this exceeds $hT$ by the (large) factor $1/(4 \pi
h^2)$. The mismatch between thermal mass and Hubble damping becomes
even worse at earlier times, since $H \propto T^4$. We thus conclude
that the condensate will not undergo early oscillations.

\section{Numerical results}
\setcounter{footnote}{0}

We are now ready to present numerical results. For the reasons given
in the previous Section, we focus on scenarios where the RH sneutrino
condensate decays through the perturbative single--particle decay
discussed in Sec.~III, i.e. we impose the lower and upper bounds
(\ref{nopreheat}) and (\ref{trmax}), respectively, on the reheat
temperature. The baryon asymmetry is computed by solving the equations
of motion (\ref{eom}) for $\tilde N$ as discussed in Sec.~II, and
using the lepton asymmetry in the sneutrino condensate obtained from
this solution via Eq.(\ref{nasym}) as input in
Eq.(\ref{eom_l}).\footnote{It is more convenient to numerically solve
the equations of motion for the ``co--moving'' fields $t N_{R,I}(t)$
as well as for the co--moving lepton density $t^2 L_D(t)$. As shown in
Sec.~II, the former can to good approximation be solved analytically
by Eq.(\ref{sol_eom}) if the Hubble--induced supersymmetry breaking is
weak, which is always true at sufficiently late time.} The numerical
solution of this equation [with initial condition $L_D(H = M_N) = 0$]
gives the lepton number density, which is translated in the final
baryon--to--entropy ratio using Eq.(\ref{final}).

\begin{figure}[htbp]
\vspace*{-5cm}
\epsfig{figure=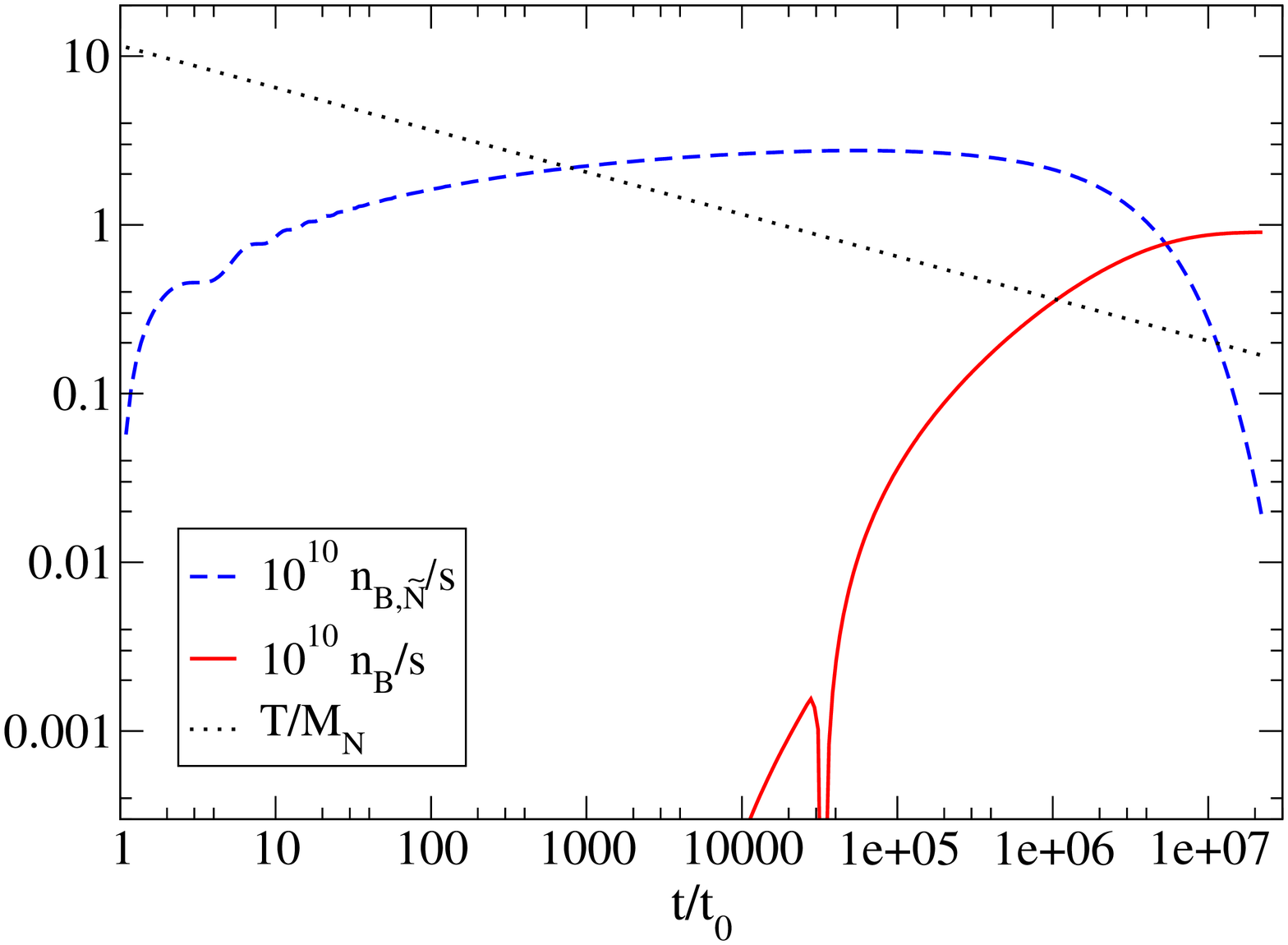,angle=-90,width=.54\textwidth,clip=}

\noindent
\epsfig{figure=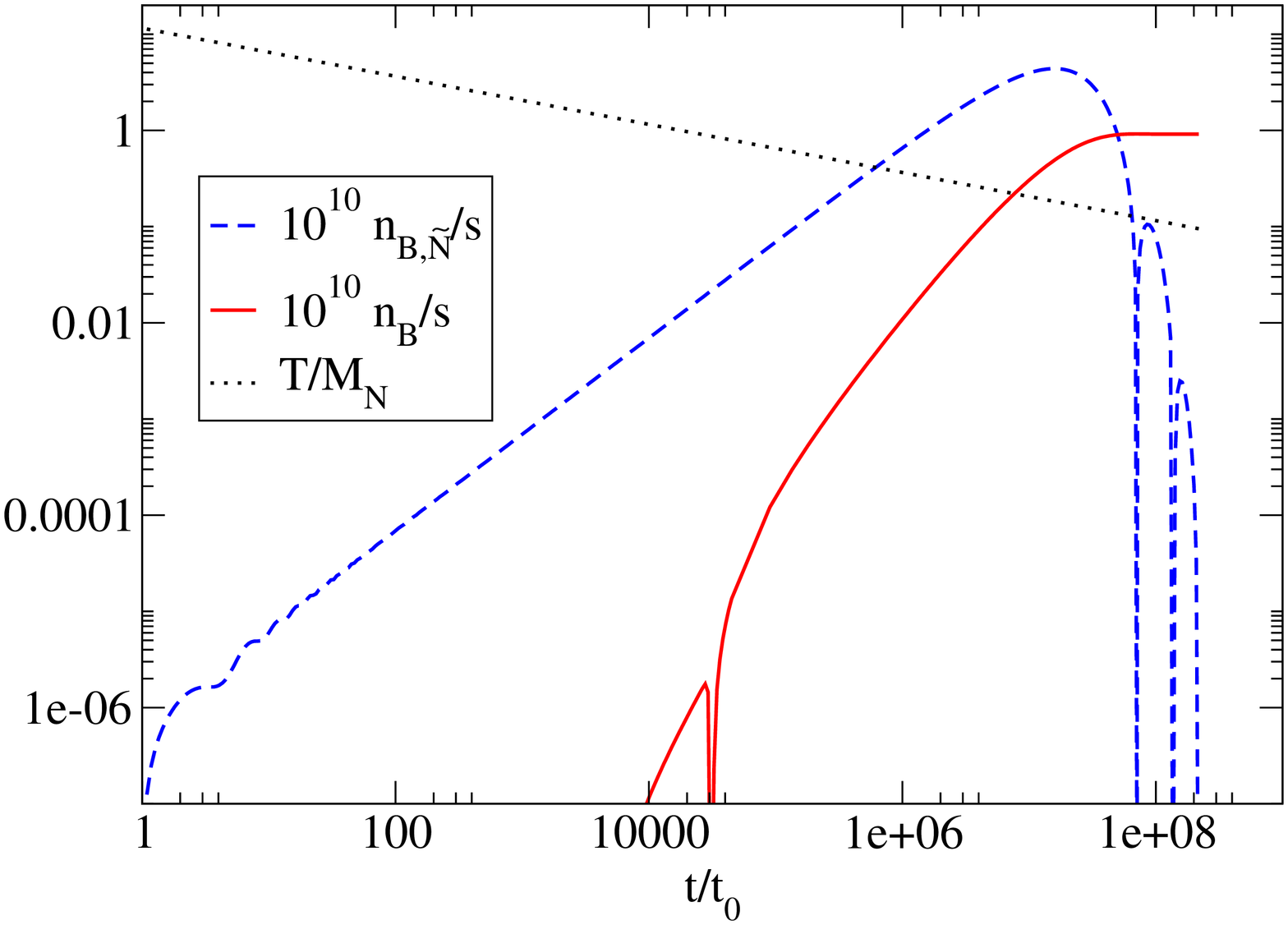,angle=-90,width=0.54\textwidth,clip=}

\caption{Time evolution of the temperature in units of $M_N$ (dotted,
black), and the absolute values of the lepton number in the RH
sneutrino condensate (dashed, blue), and in the $\tilde N$ decay
products (solid, red). The lepton numbers are normalized using
Eq.(\ref{final}), i.e. we show the baryon--to--entropy ratio that
would result if the lepton number present at time $t$ was converted
into a baryon asymmetry via sphalerons. The parameters in the upper
(lower) figure correspond to those in the first (second) row of
Table~1. We have imposed the initial condition (\ref{ini}) at $H =
M_N$, i.e. $t_0 = 2/(3M_N)$. }
\end{figure}

In Figs.~1a,b we show the evolution of the temperature and of the
lepton number stored in the $\tilde N$ condensate and its decay
products for two scenarios with rather heavy right--handed neutrinos
($M_N = 3 \cdot 10^{10}$ GeV) and high reheat temperature ($T_{\rm R}
= 10^9$ GeV). The first scenario has sizable Hubble--induced soft SUSY
breaking $b = 0.1$. As a result the lepton asymmetry
in the condensate (dashed, blue curve) grows almost monotonically with
time until $t \simeq \Gamma_N^{-1}$; the wiggles at early time are
related to the terms $\propto \Delta$ in Eq.(\ref{ln1}) (as well as to
terms not included in this approximate analytical solution). The
lepton asymmetry of the decay products (solid, red curve) remains zero
until $T = 1.2 M_N$, since for higher temperatures thermal effects
forbid perturbative $\tilde N$ decays. [Since the condition
(\ref{nopreheat}) is satisfied, non--perturbative decays of the
condensate are forbidden as well.]  Then the fermionic $\tilde N$
modes open up, leading to a negative lepton number carried by the
decay products. At $T = 0.9 M_N$ bosonic $\tilde N$ decays become
possible, and quickly take over because of the Bose enhancement. The
lepton asymmetry of the decay products therefore changes sign, and
thereafter grows monotonically. Note that $L_{\tilde N}$ remains
almost constant for quite a long time. Eventually the Hubble--induced
$B-$term becomes subdominant to the soft term, taken as $B = 100$ GeV
in this example, and $L_{\tilde N}$ starts to oscillate. However, for
the given choice $h = 0.002$ most of the condensate has already
decayed by that time, so these oscillations essentially have no effect
on the final baryon asymmetry.

The second scenario is for negligible Hubble--induced supersymmetry
breaking. Although we have increased $B$ to 1 TeV, the growth of the
asymmetry $L_{\tilde N}$ of the condensate is now much slower. Since
$T_{\rm R}$ remains the same, $\tilde N$ decays commence at the same
time as in the first scenario. Since $L_{\tilde N}$ is still quite
small at that time, the growth of $L_D$ is initially also much slower
than in the first scenario; note the different scales for the $y-$axes
of the two figures. To compensate for this, a somewhat higher maximal
value of $L_{\tilde N}$ is required; this has been obtained by
adjusting $N_0$ accordingly. In other words, scenarios with $b = 0$
are somewhat less efficient in transferring the lepton asymmetry from
the condensate to its decay products. We also note much more prominent
oscillations of $L_{\tilde N}$ than in the first case. However, the
second, ``wrong--sign'' peak in $|L_{\tilde N}|$ is suppressed by
about a factor of 50 due to $\tilde N$ decays, and can therefore not
deplete the final baryon asymmetry significantly.

Successful baryogenesis is also possible for significantly smaller
values of $M_N, \, T_{\rm R}$ and $B$. Some examples are shown in
Table~1. The first two rows are the two scenarios depicted in
Figs.~1. Reducing $M_N$ requires that the product $h T_{\rm R}$ be
reduced as well to satisfy the condition (\ref{trmax}). At the same
time we have to satisfy the lower bound (\ref{nopreheat}) on $T_{\rm
R}$. These two conditions together imply that scenarios with reduced
$M_N$ are only successful if $h$ and $T_{\rm R}$ are reduced as well,
roughly $\propto \sqrt{M_N}$ and $\propto M_N$, respectively. The
$\tilde N$ decay width then scales $\propto M_N^2$, which requires a
corresponding reduction of $|B|$ to maintain the ``matching'' $|B|
\simeq \Gamma_N$ required for an effective transfer of the asymmetry
from the condensate to the decay products. The approximate result
(\ref{final_app}) indicates that the final baryon asymmetry will be
unchanged if the ratio $M_N/T_{\rm R}$ remains constant, and indeed we
see that the required initial field value $N_0$ does not change too
much in the successful scenarios shown in the Table. Note that this
value is somewhat below the scale of Grand Unification $M_{GUT} \simeq 2
\cdot 10^{16}$ GeV. This is compatible with $N$ being charged under
the GUT gauge group, in which case the new $D-$term limits $|N_0|
\lsim M_{GUT}$ as discussed earlier.

\begin{table} \label{table1}
\caption{Some examples of parameters which lead to a successful soft
leptogenesis from the $\tilde N$ condensate for $\sin 2 \theta_0 \sim
1$ and respect the constraints (\ref{nopreheat}) and (\ref{trmax}) on
$T_{\rm R}$. All dimensionful parameters are in GeV. }
\vspace*{3mm}
\begin{tabular}{|c|c|c|c||c|c|}
$B$ & $b$ & $M_N$ & $T_{\rm R}$ & $N_0$ & $h$ \\
\hline
\hline
$10^2$ & $0.1$ & $3 \cdot 10^{10}$ & $10^9$ & $5 \cdot 10^{14}$ & 
$2 \cdot 10^{-3}$ \\
\hline
$10^3$ & $0$  & $3 \cdot 10^{10}$ & $10^{9}$ & $1.2 \cdot 10^{15}$ &
$10^{-3}$ \\ 
\hline 
$1$ & $0$ & $10^{9}$ & $3 \cdot 10^7$ & $2 \cdot 10^{15}$ & $10^{-4}$  \\
\hline
$10^{-2}$ & $0$ & $10^{8}$ & $4 \cdot 10^6$ & $1.5 \cdot 10^{15}$ & 
$3 \cdot 10^{-5}$  \\
\hline 
$10^{-4}$ & $0$ & $10^{7}$ & $10^{6}$ & $5 \cdot 10^{14}$ & $10^{-5}$ \\
\hline 
$10^{-6}$ & $0$ & $3 \cdot 10^{5}$ & $10^{4}$ & $7 \cdot 10^{14}$ & $10^{-5}$ 
\end{tabular}
\end{table}

It is also necessary that the ``inverse decay'' of $\tilde N$ does not 
erase the generated lepton asymmetry; that is, lepton number violating
processes where a (real or virtual) $\tilde N$ (or $N$) is exchanged
must not be in equilibrium. For $T \ll M_N$, these processes can only
proceed via the exchange of a far off--shell $\tilde N$ or $N$, and
occur at a rate $\propto h^4 T^3/M_N^2$. The upper bounds
(\ref{trmax}) and (\ref{nopreheat}) on $h$ imply that these reactions
will not be in equilibrium. On the other hand, Figs.~1 indicate that
the most relevant temperature for creating the lepton asymmetry in the
decay products is not too far below $M_N$; indeed, $|\Delta_{BF}|$
becomes too small for our scenario to be workable (given the other
constraints) if the condensate decays at a temperature $T \ll
M_N$. The creation of (nearly) on--shell $\tilde N$ particles from the
thermal plasma is then at most mildly Boltzmann--suppressed. However,
due to the very small width of $\tilde N$, $2 \rightarrow 1$ processes
(the $N$ inverse decay in the stricter sense) will be kinematically
possible only for a very small fraction of all collisions of light
particles in the plasma. We estimate the rate for these processes at
$T \gsim M_N$ as
\beq \label{invdec}
\Gamma_{\rm inv} \sim \frac {\Gamma_0}{4} \frac {2T} {M_N}
\int_{M_N^2/(4T^2)}^\infty \frac {x \, dx} { e^x \pm 1},
\eeq
where the $+$ ($-$) sign is for fermionic (bosonic) initial
states. Figs.~1 show that most of the asymmetry in the decay products
is created at $T \lsim M_N/3$. According to Eq.(\ref{invdec}) this
implies $\Gamma_{\rm inv} \lsim \Gamma_N/5$, which should be compared
with the Hubble expansion rate at $H \simeq \Gamma_N$. We therefore
expect at worst a small dilution of the asymmetry through inverse
$\tilde N$ decays, which can easily be compensated by a small increase
of $N_0$. Higher order ($2 \rightarrow 2$) scattering reactions
producing on--shell $\tilde N$ particles are possible for any
configuration with cms energy $\geq M_N$ (as opposed to almost exactly
equal to $M_N$ in case of $2 \rightarrow 1$ reactions), but are
suppressed by an additional coupling and additional $\pi$ factors from
phase space integration; they should thus be subdominant to inverse
$\tilde N$ decay.

We just saw that successful baryogenesis requires quite small values
of the soft breaking parameter $B$ if $M_N \lsim 10^9$ GeV. In models
of gravity--mediated supersymmetry breaking, the exact value of $B$
depends on the details of supersymmetry breaking sector and the
structure of the K\"ahler potential. For example, in the Polonyi model
and with minimal kinetic terms one obtains $B = (2 -\sqrt{3})
m_0$~\cite{nilles}. Under general circumstances one can expect that
$|B| \sim m_0$. However, for a non--minimal K\"ahler potential it is
possible to have $|B| \ll m_0$\footnote{Non--minimal K\"ahler terms
for $\tilde N$ may be preferred by other considerations. For example,
it is required for the flatness of the potential at $|\langle {\tilde
N}\rangle| > M_{\rm P}$, if $\tilde N$ is the
inflaton~\cite{sninfl}.}. Similarly, simple models of gauge--mediated
supersymmetry breaking have $B = 0$ at the messenger scale. However,
even if $B = 0$ at the tree--level, it is inevitably generated at the
one--loop level through the $A-$term associated with the neutrino
Yukawa coupling\footnote{If this $A-$term itself vanishes at
tree--level, it will be generated through gauge interactions from the
electroweak gaugino masses. If ${\bf N}$ carries some gauge charge,
$B$ will also be generated at one--loop through gauge interactions.},
and hence in general $B$ will not remain strictly zero. Note that
these loop corrections are ${\cal O}(h^2)$, so any value of $m_0 \gsim
|B| \gsim h^2 m_0$ is (technically) natural. The values of $B$ used in
Table~1 fall in this range if the scale of (most) soft breaking
parameters $m_0 \sim 1$ TeV.\footnote{If $N$ carries some gauge
charge, and the messenger scale is larger than the scale where the
corresponding gauge symmetry is broken, the natural lower bound on
$|B|$ increased to $\sim m_{1/2}/(16 \pi^2)$, where $m_{1/2} \gsim
100$ GeV is a gaugino mass; this excludes scenarios with $|B| \ll 1$
GeV.}

\begin{figure}[htbp]
\epsfig{figure=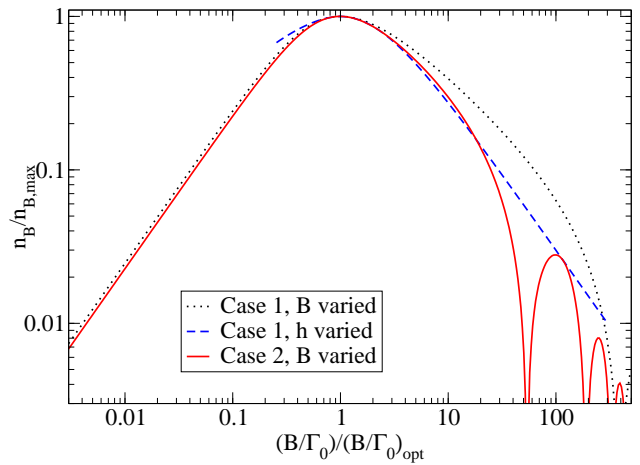,angle=-90,width=.54\textwidth,clip=}

\caption{Dependence of the final baryon asymmetry on the deviation of
the ratio $B/\Gamma_0$ from its optimal value, which is near 1;
$\Gamma_0 = h^2 M_N/(4\pi)$ is the zero--temperature width of the RH
sneutrino $\tilde N$. Dotted (black) and solid (red) curves correspond
to the second and last scenario of Table~1, except that $B$ has been
varied. The dashed (blue) curve again corresponds to the second line
of Table~1, but now the coupling $h$ has been varied.}
\end{figure}

The results of the Table confirm that Eq.(\ref{final_app}) provides a
reasonable approximation if the Hubble$-b$ can be neglected and the
resonance condition $\Gamma_N \simeq |B|$ is satisfied, with
$\Delta_{BF} \simeq 0.05$ to $0.25$. Fig.~2 shows how the final baryon
asymmetry is reduced when we deviate from this relation. We see that
the asymmetry has a clear maximum when plotted against the ratio
$B/\Gamma_0$, where $\Gamma_0$ is the width of $\tilde N$ at
temperature $T = 0$. However, the curve is rather asymmetric around
its maximum. If $|B|$ is reduced from its optimal choice, leaving $h$
(and hence $\Gamma_0$) unchanged, one finds an almost universal
behavior once both the asymmetry and the ratio $|B|/\Gamma_0$ are
normalized to their values at the maximum of $n_B/s$; for $|B|$ well
below the optimal choice, the asymmetry becomes simply $\propto
|B|$. This can be understood from the fact that the lepton number of
the condensate at time $t \simeq 1/\Gamma_N$ is $\propto |B|$ for
small $|B|$, see Eq.(\ref{nasym}).

On the other hand, different scenarios give different results when
$|B|$ is increased beyond its optimal choice. The crucial variable
here is the temperature of the plasma at time $t \simeq 1/\Gamma_N$,
divided by $M_N$. This quantity is nearly two times bigger in ``Case
2'' of Fig.~2 (corresponding to the last row of the Table) than for
``Case 1'' (the second row of the Table). A smaller temperature means
that the quantity $|\Delta_{BF}|$ will decrease faster with time, see
Eq.(\ref{deltabfapp}). Hence the final baryon asymmetry remains
dominated by the contribution from the {\em first} oscillation of the
lepton number $L_{\tilde N}$ of the condensate, even if only a rather
small fraction of the condensate decays during that time; however,
eventually the oscillations of $L_{\tilde N}$ will manifest itself in
the final baryon asymmetry as well. On the other hand, increasing the
temperature increases the importance of subsequent minima and maxima
of $L_{\tilde N}$, thereby leading to earlier oscillations of the
final baryon asymmetry (as function of $B$, not as function of time),
as shown by the solid curve.

The final asymmetry also depends on whether $|B|$ or $h$ (and hence
$\Gamma_N$) is varied. This can be seen by comparing the dashed and
dotted curves in Fig.~2. Reducing $h$ also increases $|B|/\Gamma_N$,
i.e. the condensate decays ``too late''. However, in addition to
having to average over several oscillations of $L_{\tilde N}$, a
reduction of $h$ for fixed $T_{\rm R}$ also implies that the
temperature at $t \simeq 1/\Gamma_N$ is reduced; recall that
$|\Delta_{BF}|$ depends exponentially on this ratio, see
Eq.(\ref{deltabfapp}). A reduction of $h$ therefore first leads to a
faster decrease of the baryon asymmetry than an increase of $|B|$
does; conversely, an increase of $h$ beyond the optimal choice leads
to a smaller reduction of $n_B/s$ than an increase of $|B|$
does. However, we just saw that reducing the temperature at $\tilde N$
decay also reduces the importance of the minima (and later maxima) of
$L_{\tilde N}$. One therefore does not find oscillatory behavior of
$n_B/s$ when plotted as function of $h$. A final difference is that $B$
can be varied arbitrarily without violating any consistency
conditions. In contrast, for fixed values of the other parameters, $h$
is bounded from above by the requirement (\ref{trmax}) that the
condensate should not be destroyed by scattering off the thermal
plasma, as well as by the requirement (\ref{nopreheat}) that
non--perturbative $\tilde N$ decays be shut off thermally; the latter
requirement is the more stringent one in Fig.~2. $h$ is also bounded
from below, by the requirement that $\tilde N$ decays before the
inflaton does.

\section{Discussion}

In this section we discuss the merits and consequences of our
scenario, compare it with similar suggestions in the literature, and
discuss what happens when some of our assumptions are relaxed.

\subsection{Consequences for low--energy physics}

The main particle physics motivation for considering heavy RH
neutrinos is undoubtedly the see--saw mechanism \cite{seesaw}, where
the masses of the light neutrinos come out quadratic in their Yukawa
coupling, and inverse to the masses of the heavy neutrinos. Our
mechanism will work for a single RH neutrino superfield, which may not
be the lightest one. Therefore the effective neutrino mass\footnote{In
our scenario the coupling $h$ enters only via the total decay width of
$\tilde N$. The existence of three light generations of (s)leptons can
therefore trivially be incorporated by the replacement $h^2
\rightarrow \sum_{i=1}^3 h_i^2$.} $\overline m = h^2 v_u^2/M_N$, where
$v_u = \langle H_U^0 \rangle \sim 150$ GeV, is even more difficult to
interpret than in usual leptogenesis models, where only the lightest
RH (s)neutrino is relevant. It nevertheless seems reasonable to
require
\beq \label{hmax}
\overline{m} \lsim 1 \ {\rm eV} \Longrightarrow h \lsim 10^{-2}
\sqrt{\frac {M_N} {10^9 \ {\rm GeV}}}\,.
\eeq
Otherwise realistic neutrino mass spectra could at best be obtained at
the cost of severe cancellations. Note that the entries of Table~1
correspond to significantly smaller values, $\overline{m} \lsim 0.001$
eV. Atmospheric neutrino oscillations indicate that at least one
neutrino must have a mass $\gsim 0.05$ eV \cite{neu_osc}. This
indicates that the other RH neutrinos must play an important role in
the see--saw mechanism, even though they may be irrelevant for
baryogenesis. Since our mechanism is not sensitive to flavor mixing,
we cannot sharpen the predictions for (s)lepton flavor violating
processes that might be accessible at low energies, compared to
``generic'' see--saw models \cite{lfv}. Moreover, in our scenario CP
is broken purely spontaneously in the early Universe, through the
complex vacuum expectation value of $\tilde N$. This obviates the need
for any complex phases in the Lagrangian. Hence no prediction can be
made for possible CP violation in neutrino oscillations \cite{cp_osc}.

\subsection{Comparison with related scenarios}

The fact that our mechanism works for purely real Lagrangian
distinguishes it from the proposal of Refs.~\cite{nir,agr}, in which
the lepton asymmetry is also created with the help of the $B-$term but
no condensate exists. There the $\tilde N-{\tilde N}^*$ asymmetry
arises from an absorptive part in the ${\tilde N}-{\tilde N}^*$
mixing, and is proportional to the imaginary part of the $A-$term
associated with the neutrino Yukawa coupling, denoted by $A_\nu$. This
results in~\cite{agr}
\beq \label{giudice}
{L_D \over (n_{\tilde N} + n_{\bar {\tilde N}})} \simeq {4 |B| \Gamma_N 
\over 4 |B|^2+ \Gamma^2_N} {\Im {\rm m} A_\nu \over M_N} \Delta_{BF}.
\eeq
A necessary condition for generating a sufficiently large asymmetry is
therefore that the phase of $A_\nu$ is large (in the basis where $B$
is real). On the other hand, constraints on the electric dipole
moments of the electron and neutron imply that the phases of many
other soft breaking terms must be small \cite{smallphase}. This may
complicate the construction of complete models where the mechanism of
refs.\cite{nir,agr} is workable. In our scenario the ratio of
Eq.(\ref{giudice}) is simply given by $\Delta_{BF}$ if $\Gamma_N$ and
$|B|$ match perfectly; the additional suppression factor due to
possible mismatch has been shown in Fig.~2.

If $\tilde N$ has to be produced thermally, the case primarily studied
in~\cite{nir,agr,gnrrs}, the asymmetry in Eq.~(\ref{giudice}) will be
too small when $|A_\nu| \simeq |B| \sim 1$ TeV. For $M_N \leq 10^9$
GeV (to be compatible with the gravitino bound on $T_{\rm R}$) the
condition for out--of--equilibrium decay of the sneutrino $\Gamma_N <
M^2_N/M_{\rm P}$ implies that $\Gamma_N < 1$ GeV. Thus the first two
terms on the RH side of~(\ref{giudice}) combine to give a number $<
M_N/M_{\rm P}$. The resultant asymmetry, after taking $\Delta_{BF}$
and the entropy factor into account, will be $\ll 10^{-10}$. In fact,
it is found in~\cite{agr,gnrrs} that the thermal scenario works only
if $|B| \ll 1$ TeV, for $10^6 \ {\rm GeV} \lsim M_N \lsim 2 \cdot
10^8$ GeV.

The situation is somewhat better for non--thermal scenarios. For
example, consider the case when the (perturbative) inflaton decay to
the sneutrino and the subsequent decay of $\tilde N$ to light fields
reheats the Universe and generates a lepton asymmetry. In this case
the $\tilde N + \tilde N^*$ number density (after inflaton decay) is
essentially equal to the inflaton number density (before decay). Using
Eqs.(\ref{giudice}) and (\ref{iratio}) we find
\beq \label{case1}
{L_D \over s} \sim {4 |B| \Gamma_N \over 4 |B|^2+ \Gamma^2_N}
\cdot {\Im {\rm m} [A_\nu] \over M_N} \Delta_{BF} \cdot {T_{\rm R}
\over m_\phi},
\eeq
where $m_\phi > M_N$ is the inflaton mass. The product of the first
two factors in Eq.(\ref{case1}) is nearly independent of $M_N$ so long
as $|B| \gsim \Gamma_N$, since $\Gamma_N \propto M_N$ for fixed
$h$. On the other hand, the see--saw constraint (\ref{hmax}) implies
that the maximal value of $\Gamma_N / M_N$ grows $\propto M_N$. For
$|B| \simeq \Im {\rm m} [A_\nu] \geq \Gamma_N$ Eq.(\ref{case1}) thus
implies
\beq \label{agrbound}
{L_D \over s} \sim 10^{-5} \Delta_{BF} {\overline{m} \over 1 \ {\rm eV}}
\frac{M_N} {10^9 \ {\rm GeV}} \frac {T_{\rm R}} {m_\phi}\,.
\eeq
This may allow $M_N$ down to $10^7$ GeV if $T_{\rm R} < M_N < m_\phi$,
and if the bound (\ref{hmax}) is nearly saturated. If $M_N > 10^9$
GeV, the bound (\ref{hmax}) allows $\Gamma_N > 1$ TeV. However,
Eq.(\ref{case1}) shows that such scenarios, with $\Gamma_N > |B|$ and
$M_N > 10^9$ GeV, will generally not yield a sufficient baryon
asymmetry.  On the other hand, values of $M_N$ up to $10^{10}$ GeV at
least can be accommodated if $\Gamma_N \simeq |B|$ and $\Im {\rm
m}[A_\nu] \simeq 1$ TeV.

There is yet another constraint on this scenario. Here $\tilde N$
particles are not produced at rest, so that their decay is delayed by
a time dilatation factor $\sim m_\phi / M_N$. For given $\Gamma_N$ and
$T_{\rm R}$ this would reduce the temperature at the time of $\tilde
N$ decay, leading to a large reduction of $|\Delta_{FB}|$, see
Eq.(\ref{deltabfapp}). Raising $T_{\rm R}$ is not an option, since
then inverse $\tilde N$ decay reactions would become much more
efficient, washing out any asymmetry. This scenario therefore requires
the inflaton mass to be only a factor of a few larger than $M_N$.

Table~1 shows that the value of $M_N$ is less constrained in our
scenario, and works for any inflaton mass $>M_N$. Here sneutrinos
exist in the form of a condensate and there is no need to produce them
after inflation. On the other hand, the constraints (\ref{trmax}) and
(\ref{nopreheat}) require rather small values of $\overline m$ in our
scenario, as remarked above.

\subsection{Additional thermal effects}

Note first that the existence of a thermal bath at $H > \Gamma_\phi$,
with temperature $T > T_{\rm R}$, does not contradict the gravitino
bound. The reason is that, for $\Gamma_N > \Gamma_\phi$, the Universe
will become RD only later when reheating completes. This stage
releases enough entropy to dilute any gravitinos produced at $T >
T_{\rm R}$~\cite{grt}.

The last rows in Table~1 have quite low values of $|B|$, well below 1
GeV. Such values may be natural in models of gauge--mediated
supersymmetry breaking, where one might expect the gravitino mass to
be near $|B|$. In this case the scaled relic density of gravitinos is given
by~\cite{mmy,dmm}:
\beq \label{gauge}
\Omega_{3/2} h^2 \simeq 0.8 \left({M_3 \over 1~{\rm TeV}}\right)^2 
\left({10~{\rm MeV} \over m_{3/2}}\right) \left({T_{\rm R} \over 
10^{6}~{\rm GeV}}\right),
\eeq
where $h \simeq 0.7$ is the present scaled Hubble constant, and $M_3$
is the gluino mass. Then, for $M_3 \sim 500$ GeV, the dark matter
limit leads to the constraint $T_{\rm R} \leq 10^8 m_{3/2}$. We thus
see that we need $m_{3/2} \gg |B|$ in the last two examples shown in
Table~1, i.e. we have to require that the leading (in inverse powers
of $M_{\rm P}$) operator that could give rise to a $B-$term is
suppressed.

Note that the instantaneous thermal bath may well have a temperature
$\gg M_N$ at early times, and can in principle produce $\tilde N$ and
${\tilde N}^*$ quanta with a number density $n_{\tilde N,th} \simeq
0.2 T^3$. This may be larger than the number density of the zero--mode
$\tilde N$ quanta in the condensate $n_{\tilde N,con} \simeq H^2
N^2_0/M_N$ (which is typically much smaller than $T^3$). One might
therefore wonder whether the former could yield a lepton asymmetry,
according to~\cite{nir,agr,gnrrs}, dominating that generated by the
latter.  Eq.~(\ref{giudice}) shows that this will not be the case so
long as $n_{\tilde N, th}/n_{\tilde N, con}< M_N/|A_\nu|$ when $H
\simeq |B|$ (note that at the time of sneutrino decay $n_{\tilde N,
th} < T^3$).  Indeed, for all the examples in Table~1, the asymmetry
generated by the condensate is dominant.

\subsection{Effects of additional RH (s)neutrinos}

So far we have considered only one of the sneutrinos. Indeed, one of
the merits of this scenario is that it works well with just one
generation of RH (s)neutrinos. Nevertheless, the other sneutrinos can
also acquire a VEV during inflation and their post--inflationary
dynamics may contribute to and/or affect the generated baryon
asymmetry. Let us denote the three sneutrinos by ${\tilde N}_1$,
${\tilde N}_2$, ${\tilde N}_3$ with masses $M_1 < M_2 < M_3$
respectively. If $M_i > H_I$, ${\tilde N}_i$ will not develop a VEV
thus being irrelevant for the production of the asymmetry. If $H_I
\gsim 10^{13}$ GeV as in simple models of inflation \cite{infl}, these
(s)neutrinos will then also be too heavy to wash out the asymmetry.

It is conceivable that all sneutrino fields whose mass $M_i < H_I$
have the same value $N_0$ (up to fluctuations of size $H_I \ll N_0$)
after inflation\footnote{This, for example, naturally happens if the
minimum of the sneutrino potential is set by a negative
Hubble--induced soft ${\rm mass}^2$ and $D$--terms under which the
sneutrinos are non--singlet.}; these field values will remain
essentially constant until $H = M_i$, at which point the amplitude of
$\tilde N_i$ begins to be redshifted $\propto 1/t$. The number density
of ${\tilde N}_i$ at $H \simeq M_1$ will therefore be $n_i =
(M^2_1/M_i) N^2_0$. In this case ${\tilde N}_1$ would be the best
candidate to yield the maximum lepton asymmetry. Then, however,
${\tilde N}_2$ and ${\tilde N}_3$ should not decay too fast such that
the resulting thermal effects destroy the coherence of ${\tilde
N}_1$. On the other hand, if the asymmetry is generated from ${\tilde
N}_3$, then ${\tilde N}_1$ and ${\tilde N}_2$ should decay before
dominating the Universe. Otherwise, such a late decay will dilute the
asymmetry. An efficient leptogenesis from ${\tilde N}_2$ requires that
${\tilde N}_3$ decay not destroy the $\tilde N_1$ condensate and
${\tilde N}_1$ decay not dilute the generated asymmetry.

Finally, independently of whether lighter RH sneutrinos (if any) have
a large VEV during inflation, they could dilute the asymmetry through
scattering reactions, analogous to the discussion of ``inverse
decays'' of the sneutrino that produced the asymmetry in the first
place. This discussion shows that the most dangerous situation occurs
when the temperature is near the mass of these lighter sneutrinos. If
this mass is less than the reheat temperature, it follows that the
effective mass $\overline{m}_{\rm light}$ associated with these light
sneutrinos should be below $\sim 0.1$ eV; not surprisingly, this is
essentially the same bound one derives from analogous arguments in
standard thermal leptogenesis \cite{plumacher}. This bound can be
relaxed by an order of magnitude or so of the lighter neutrino masses
lie above $T_{\rm R}$, thereby allowing to saturate the
phenomenological bound (\ref{hmax}), which of course applies to
the effective neutrino masses associated with any of the RH neutrino
fields. 

We conclude that our scenario works most comfortably when the
dynamics of the other sneutrinos does not affect the picture presented
in previous sections. Then the generated baryon asymmetry will be
given by the expression in~(\ref{final_app}), with $\tilde N$ being
the sneutrino which results in the maximum asymmetry. The decay or
exchange of other sneutrinos can suppress the produced asymmetry, thus
constraining the scenario.

\subsection{Late decaying sneutrino}
\setcounter{footnote}{0}

So far we have assumed that the sneutrino decays before the inflaton
does. If $\Gamma_N < \Gamma_\phi$ our scenario will not be altered
qualitatively, so long as $\tilde N$ does not dominate the energy
density of the Universe (see the next Subsection). Since the
co--moving entropy density after inflaton decay is essentially
constant in this case, the dilution factor (\ref{iratio}) is the same
as for $\Gamma_N > \Gamma_\phi$. The main difference is that the
temperature decreases much faster in the radiation--dominated era: $T
\propto t^{-1/2}$, as compared to $T \propto t^{-1/4}$ during the
matter--dominated era after inflation. This means that there is a
relatively shorter time window where $T \sim M_N$, which yields the
maximal $|\Delta_{BF}|$. As a result, the effective $|\Delta_{BF}|$
will be slightly smaller than in scenarios with $\Gamma_N >
\Gamma_\phi.$ Moreover, for given temperature\footnote{Recall that in
the matter--dominated era the hot plasma contributes only a small
fraction to the total energy density.} the Hubble parameter in the
matter--dominated epoch exceeds that in the radiation--dominated era
by a factor $(T/T_{\rm R})^2$. Wash--out reactions are therefore more
dangerous if $\Gamma_N < \Gamma_\phi$.

\subsection{${\tilde N}$ as the inflaton or curvaton}

So far our estimates of the baryon asymmetry were based on the
assumption that $\tilde N$ decays before dominating the Universe. In
fact, when $|B| \simeq m_0 = 1$ TeV, this will be necessary to avoid
gravitino overproduction. However, $\tilde N$ domination is possible
if the sneutrino is the inflaton, or the
curvaton~\cite{curvaton,mt}. Then the maximum lepton asymmetry will be
given by
\beq \label{dom}
{L_D \over s} \simeq {3 T_{{\rm R}} \over 4 M_N} \Delta_{BF},
\eeq
where $\Delta_{BF}$ is given by~(\ref{deltabf}) at $T=T_{\rm R}$. Here
$T_{\rm R} \simeq 1.4 \left( g_*^{-1/2} \Gamma_N M_{\rm Pl}
\right)^{1/2}$ is the reheat temperature of the Universe after the
decay of $\tilde N$. Note that for $T_{\rm R} < M_N/30$, we simply
have $|\Delta_{BF}| \approx (T_{\rm R}/M_N)^2$, see
Eq.~(\ref{deltabfapp}).

If $\tilde N$ is the inflaton~\cite{sninfl,ery}, $M_N = 10^{13}$ GeV
(to have density perturbations of the correct size) and, due to the
gravitino bound $T_{\rm R} \leq 10^{9}$ GeV, $\Gamma_N \leq 10$ GeV.
Therefore obtaining the maximum asymmetry requires that $|B| \leq 10$
GeV.  Even then, the asymmetry will be too small by two orders of
magnitude. A sufficient asymmetry can be generated with a higher
$T_{\rm R}$ at the expense of gravitino overproduction. Then, however,
late entropy production will be necessary in order to dilute the
excess of gravitinos. Note that $(L_D/s) \propto T^3_{\rm R}$, see
Eq.~(\ref{dom}), while $(n_{3/2}/s) \propto T_{\rm
R}$~\cite{ellis}. Entropy release can therefore dilute gravitinos
while yielding an acceptable asymmetry. For example, if $\Gamma_N =
|B| = 1$ TeV and $T_{\rm R} = 10^{11}$ GeV, successful leptogenesis
will require a dilution factor of $10^4$.

If $\tilde N$ is the curvaton~\cite{hmy,mm,m}, $M_N$ and $T_{\rm R}$ can
be much smaller. Then a sufficient asymmetry can be generated, provided
that $10^3 T_{\rm R} \geq M_N$, where $T_{\rm R}$ is the temperature
of the Universe after $\tilde N$ decay. In this case the sneutrino
dominance typically requires a tiny $h$ and a $N_0$ not much smaller than
$M_{\rm P}$. This implies that $\Gamma_N \ll m_0$, and hence $|B| 
\ll m_0$ is required to have the maximum asymmetry in the condensate. A
larger $|B|$ can be compensated for by appropriate increase in $T_{\rm
R}/M_N$.

\section{Summary and conclusions}

In this paper we have discussed a new variant of the leptogenesis
mechanism for the creation of the baryon number of the Universe. Our
starting point is the observation that the complex scalar field
describing the superpartners of $SU(2) \times U(1)_Y$ singlet
(``right--handed'') neutrinos, introduced to explain the smallness of
the observed neutrino masses through the see--saw mechanism, should
acquire large expectation values during inflation, if their mass is
less than the Hubble scale during inflation. Note that this argument
applies to the real and imaginary parts of this field, i.e. one
expects CP to be broken spontaneously in such a scenario. In addition,
lepton number is violated in the scalar sector by the soft
supersymmetry breaking $B-$term associated with the (supersymmetric)
Majorana mass term for the heavy neutrinos. The third Sakharov
condition is automatically satisfied, since the creation and decay of
such a condensate is a non--equilibrium process.

We showed that such a scenario can indeed create the required baryon
asymmetry if a number of conditions are satisfied. To begin with, the
$B-$ term creates an oscillating lepton number in the condensate. This
can be transferred efficiently to light (s)particles only if the
period of this oscillation, given by $|B|$, is not too different from
the decay width $\Gamma_N$ of the heavy sneutrino. Moreover, this
transfer requires supersymmetry to be broken, the dominant
contribution coming from thermal effects. This requires that the
re--heat temperature $T_{\rm R}$ should not be too much below the mass
$M_N$ of the heavy (s)neutrinos. Moreover, the condition that the
sneutrino condensate should not be destroyed by thermal effects
imposes an upper bound (\ref{trmax}) on $T_{\rm R}$. On the other
hand, the lower bound (\ref{nopreheat}) on $T_{\rm R}$ can be derived
if one requires that thermal effects should shut off possible
non--perturbative $\tilde N$ decay mechanisms. As discussed in
Sec.~IV~A, this may not be necessary, since other effects may prevent
too early non--perturbative decays of the $\tilde N$
condensate. Recall also that non--perturbative effects might even {\em
increase} the efficiency of our mechanism. However, in the absence of
a complete quantitative understanding of these non--perturbative
effects, we focused on scenarios which can be described purely
perturbatively. In that case the condition of sufficient baryogenesis
alone imposes both upper and lower bounds on $T_{\rm R}$, both of
which scale $\propto M_N$ as discussed in Sec.~V. This is to be
contrasted with thermal leptogenesis, which only leads to a lower
bound on $T_{\rm R}$ -- which, however, is uncomfortably close to the
upper bound on $T_{\rm R}$ from gravitino overproduction.

Note that $T_{\rm R} < M_N$ is required if $\tilde N$ is to decay
before the inflaton does; we saw in Sec.~VI E that in the opposite
situation our mechanism will be somewhat less efficient. Our scenario
therefore prefers $T_{\rm R} \lsim 0.1 M_N$, whereas conventional
thermal leptogenesis works best if $T_{\rm R} \gsim M_N$. Given the
bound $T_{\rm R} \lsim 10^9$ to avoid overproduction of gravitinos,
our mechanism can accommodate significantly larger values of $M_N$, up
to a few times $10^{10}$ GeV; however, again in contrast to thermal
leptogenesis, it can work also with much smaller $M_N$, so long as
$|B|$ and the Yukawa coupling $h$ are reduced as well. On the other
hand, if non--perturbative $\tilde N$ decays are shut off by thermal
effects, our scenario requires the effective light neutrino mass
associated with this particular heavy (s)neutrino to lie in a rather
narrow range around 1 meV. The smallness of the Yukawa coupling also
means that the heavy (s)neutrino superfield responsible for
leptogenesis in this mechanism does not leave any imprint on the
low--energy sparticle spectrum, and will therefore not contribute at
any appreciable level to slepton--mediated lepton flavor violating
processes. However, since our scenario works for a single heavy
(s)neutrino (which may only couple to a single light neutrino), little
can be said about the contributions of the other heavy neutrinos to
the see--saw mass matrix of the light neutrinos. This should allow to
construct models that incorporate our mechanism and yield a realistic
mass matrix for the light neutrinos.

The main numerical results of our paper are Eq.(\ref{final_app}),
which adequately describes the final baryon asymmetry if the $\tilde
N$ decay width perfectly matches the frequency with
which the lepton number of the $\tilde N$ condensate oscillates, and
Fig.~2, which shows how a mismatch between these two quantities
reduces the baryon asymmetry. This last figure is qualitatively
similar to the dependence of $n_B$ on the effective light neutrino
mass in models of thermal leptogenesis.

The perhaps least appealing aspect of this mechanism is that the final
result depends (quadratically) on the absolute value $N_0$ of the
field at the end of inflation, and (linearly) on the $\sin$ of its
phase. The latter is naturally expected to be ${\cal O}(1)$, while the
former could easily be sufficiently large for our purposes. On the
other hand, it might be meaningful that the model is sufficiently
constrained that it does not allow to generate a baryon asymmetry orders of 
magnitude larger than the observed one. We therefore
conclude that this mechanism should be regarded on a par with other
mechanisms for leptogenesis that have been suggested in the
literature.

\section*{Acknowledgements}

The authors wish to thank S. Davidson, A. Mazumdar, G. Moore, Y. Nir and 
M. Peloso for useful discussions and comments. The work of R.A. is supported 
by the National Sciences and Engineering Research Council of Canada. The 
research of M.D. is supported by ``Sonderforschungsbereich 375 f\"ur 
Astro--Teilchenphysik'' der Deutschen Forschungsgemeinschaft.


\end{document}